\documentclass[aps,pra,nofootinbib,preprint,amsmath,amssymb,floatfix,notitlepage,superscriptaddress]{revtex4-2}
\usepackage{comment}
\usepackage{mathtools}
\usepackage{graphicx}
\usepackage{hyperref}
\usepackage{amsmath, amsfonts, amssymb}

\DeclareMathOperator{\sgn}{sgn}
\DeclareMathOperator{\curl}{curl}

\usepackage{siunitx}

\usepackage{color,soul}
\begin{document}

	\newcommand{\tc}{\textcolor}
	\title{Axion electrodynamics: Green's functions, zero-point energy and optical activity}
	\author{Amedeo M. Favitta}
	\affiliation{Dipartimento di Fisica e Chimica—Emilio Segr\'{e}, Università degli Studi di Palermo, Via Archirafi 36, I-90123 Palermo, Italy }
	\author{Iver H. Brevik}      
	\affiliation{Department of Energy and Process Engineering, Norwegian University of Science and Technology, N-7491 Trondheim, Norway}
	\author{Moshe M. Chaichian}
	\affiliation{Department of Physics, University of Helsinki and Helsinki Institute of Physics, P.O. Box 64, FI-00014 University of Helsinki, Finland}

	\date{15/06/2023 \\ To To appear in the Annals of Physics}          

\begin{abstract}
Starting from the theory of Axion Electrodynamics, we work out the axionic modifications to the electromagnetic Casimir energy using the Green's function method, both when the axion field is initially assumed purely time-dependent and when the axion field configuration is a static domain wall, so purely space-dependent. For the first case it means that the oscillating axion background is taken to resemble an axion fluid at rest in a conventional Casimir setup with two infinite parallel conducting plates, while in the second case we evaluate the radiation pressure acting on an axion domain wall. We extend previous theories in order to include finite temperatures. Various applications are discussed. {\it i)} We review the theory of Axion Electrodynamics and particularly the energy-momentum conservation in a linear dielectric and magnetic material. We treat this last aspect by extending former results by Brevik and Chaichian (2022) and Patkos (2022).
{\it ii)} Adopting the model of the oscillating axion background we discuss the axion-induced modifications to the Casimir force between two parallel plates by using the Green's function method.
{\it iii)} We calculate the radiation pressure acting on an axion domain wall at finite temperature $T$.
Our results for an oscillating axion field and a domain wall are also useful for condensed matter physics, where some topological materials, "axionic topological insulators", interact with the electromagnetic field with a Chern-Simons interaction, like the one in Axion Electrodynamics, and there are experimental systems analogous to time-dependent axion fields and domain walls as the ones showed e.g. by Jiang, Q. D., \& Wilczek, F. (2019) and Fukushima et al. (2019). {\it iv)} We compare our results, where we assume time-dependent or space-dependent axion configurations, with the discussion of the optical activity of Axion Electrodynamics by Sikivie (2021) and Carrol et al. (1990). We also make comparisons with the properties of known materials, such as optically active, chiral media and the Faraday effect.
\end{abstract}
\maketitle
\bigskip

\section{Introduction}\label{secintro}
Axions, the hypothetical particles suggested  by Peccei and Quinn \cite{PhysRevLett.38.1440,Peccei2008,https://doi.org/10.48550/arxiv.2105.01406} in order to solve the strong CP problem in QCD \cite{bigi2021new,PhysRevLett.40.223,schwartz2014quantum}, have attracted a strong theoretical interest  because they are among some of the best known candidates for dark matter, as first suggested in Refs.\cite{Preskill:1982cy,Abbott:1982af,Dine:1982ah}. This topic is one of the most important problems in cosmology, and is discussed lively in the current literature \cite{ amruth2023einstein, Peccei2008,Sikivie2008, PhysRevD.85.105020,PhysRevD.91.065014,RevModPhys.93.015004,Millar_2017,doi:10.1142/S0217751X22501512}.
The axion is still a hypothetical and undetected particle and there have been many efforts to detect it.
A most promising way to detect axions is based on its interaction with the electromagnetic field.
Axion Electrodynamics is a natural extension of classical electrodynamics, where the axion field is coupled to the electromagnetic field, and this interaction between the photonic field and the axion field makes it possible to detect the axions experimentally  \cite{RevModPhys.93.015004,app12136492}.
It is worth noticing  that the interaction between the  electromagnetic field and the axion field leads to modifications of the dispersion relation and the zero-point energy of the electromagnetic field, similarly to chiral QED, another interesting extension of quantum electrodynamics that is investigated in Refs.~\cite{canfora2022casimir,oosthuyse2023interplay}.
These last aspects are discussed in our work along with the calculation of the axionic modified Green's functions, where we find new results, and compare them to the results of Ref.~\cite{PhysRevD.102.123011}.
This work is also related to earlier investigations in Refs. \cite{Sikivie2008, PhysRevD.32.1560, Blasi_2023, doi:10.1142/S0217751X22501512, PhysRevD.100.045013}.

We present the basics of Axion Electrodynamics and introduce our notation in Section~\ref{intro}.
Energy-momentum balance is considered in Section~\ref{tensor}.
We develop the basics of the Green's function method in Axion Electrodynamics in Section~\ref{vacgreen} and use it in practice for time-dependent axion backgrounds, namely with constant time derivative and oscillating behaviour in Section~\ref{time-dependent}. There we calculate the Casimir force between two perfectly conducting parallel plates in two time-dependent backgrounds.
We treat space-dependent axion backgrounds, namely dependent on just the space coordinate $z$, in Sections~\ref{wavedomain} and \ref{spazio}.
In Section~\ref{wavedomain} we develop the optical properties of a toy model of the axion domain wall. The electromagnetic dispersion relations in such an axion background are given in Subsection~\ref{Casimirdisperdo}.
General features of the Green's function method for domain walls are introduced in Section~\ref{spazio}.

The Casimir force on a domain wall is obtained in Subsection~\ref{Casimir-dm}. Subsection~\ref{dio} also deals with the calculation of the temperature-dependent electromagnetic radiation pressure acting on axion domain walls and we exploit our results for applications in Axion Cosmology and Condensed Matter Physics.
Section~\ref{optical} compares our results with formerly known phenomena in Classical Electrodynamics, such as the Faraday effect and optical activity in chiral media, and former results on the optical activity in Axion Electrodynamics.
Our work on the Green's functions for the two given types of axion backgrounds of experimental interest are relevant for observing an axion field in the current Universe.
This is seen in the long wavelength approximation, i.e. we assume the experimental set-up to have spatial dimension $L$ much smaller that the de Broglie wavelength of the local axion field, so that we take $a(t)=a_0 \sin(\omega_a t)$ . If a further approximation of taking the time-derivative of the axion field to be constant is just a pedagogical first approximation for QCD axion, it is exact for the axion field in Weyl semimetals.
The axion field is mostly treated as a fixed background field, since this is true at first order in the perturbation theory because there is no back reaction of the electromagnetic ﬁeld onto axions in such an approximation, partly analogous to what happen in the linear approximation for weak gravitational perturbations in General Relativity \cite{Kiefer2009}. We discuss this point in Section~\ref{tensor}, where we also partly discuss the axion dynamics when interacting with the electromagnetic field.
\section{Basic elements of Axion Electrodynamics}\label{intro}
\subsection{Lagrangian density of Axion Electrodynamics}
In the following we adopt the metric convention $\eta_{\alpha \beta}={\rm diag}(+1,-1,-1,-1)$ and the measure system of natural units system such that: $\hbar = c =4 \pi \epsilon_0 = 1$. The remaining unit is choosen to be energy, measured in \SI{}{eV}.
We consider a pseudoscalar axion field $a= a({\bf r}, t)=a(x)$ present in the entire Universe, having a two-photon interaction with the electromagnetic field.
The total Lagrangian density $\mathcal{L}$ describing the interaction between electromagnetic field and axion field in the Minkowski space inside a linear dielectric and magnetic material with dielectric permittivity $\varepsilon$ and magnetic permeability $\mu$ is \cite{doi:10.1142/S0217751X22501512, RevModPhys.93.015004, Sikivie2008, PhysRevD.85.105020,Millar_2017}:
\begin{equation} \label{eq0}
	\mathcal{L}= -\frac{1}{4} F^{\alpha \beta}H_{\alpha \beta}+\mathcal{L}_a-J^{\mu} A_{\mu}+ \frac{1}{4}g_{a \gamma \gamma} a(x) F_{\mu \nu} \tilde{F}^{\mu \nu},
\end{equation}
$\tilde{F}^{\alpha \beta}=\frac{1}{2} \epsilon^{\alpha \beta \gamma \delta} F_{\gamma \delta} $ is the dual  Faraday tensor , $\epsilon^{\alpha \beta \gamma \delta}$ is the totally antisymmetric symbol with $\epsilon^{0123}=1$, and $m_a$ is the mass of the axion. The tensor $H$ is:
\begin{equation}\label{diofa}
	H_{\alpha \beta}=
	\begin{pmatrix}
		0 & -D_x & -D_y & -D_z \\
		D_x & 0 & H_z & -H_y \\
		D_y &-H_z & 0 & H_x \\
		D_z & H_y& -H_x & 0\\
	\end{pmatrix},
\end{equation}
where $\vec{D}$ and $\vec{H}$ are the usual displacement electric and magnetic fields that are related to electric $\vec{E}$ and magnetic $\vec{B}$ fields by $\vec{D}=\varepsilon \vec{E}$ and $ \vec{B}=\mu \vec{H}$ for a linear dielectric and magnetic material. $J^{\mu}$ is an external classical electrical 4-current that we present here for purpose of completeness. Finally, $g_{a \gamma \gamma}=g_{\gamma} \frac{\alpha}{\pi} \frac{1}{f_a}  $ where $\alpha$  is the usual fine
structure constant and $f_{a}$ is the axion decay constant whose value is only insufficiently
known.  Astrophysical and experimental bounds indicate that $f_{a} \sim 10^{9}-10^{12} \,\SI{}{GeV}$ \cite{RevModPhys.93.015004}.

In most cases in the following,  we will use the non-dimensional quantities $\Theta(x)=g_{a \gamma \gamma} a(x)$ and $\theta(x)=a(x)/f_a$.
The axion mass  is related to the decay constant $f_a$ by \cite{PhysRevLett.40.223}
\begin{equation}
	m_a \simeq \frac{\sqrt{m_u m_d}}{m_u+m_d} \frac{f_{\pi} m_{\pi}}{f_a} \simeq 6 \times 10^{-6} \; \SI{}{eV} \left(\frac{10^{12} \; \SI{}{GeV}}{f_a} \right),
\end{equation}
where $m_{\pi} \simeq \SI{140}{MeV}$ is the pion mass and $f_{\pi} \simeq \SI{93}{MeV} $ is the pion decay constant. $g_{\gamma}$ is a model-dependent constant of order unity \cite{RevModPhys.93.015004}.
\subsection{QCD Axion Lagrangian, domain walls and long wavelength approximation}
$\mathcal{L}_a$ is the "free Lagrangian" of the axion field, that in the case of High Energy Physics keeps care of  the effective interaction with gluonic fields that gives rise to its mass. Here we neglect the interaction with the gravitational field. For our purposes  it is \cite{PhysRevD.85.105020}:
\begin{equation}
		\mathcal{L}_a=\frac{1}{2} \partial_{\mu} a \,\partial^{\mu} a-V_{QCD}(a,T).
\end{equation}
The form of the potential $V_{QCD}(a,T)$ is a nontrivial point   for finite  temperatures $ T$ \cite{PhysRevD.85.105020}.
Using the dilute gas approximation one may adopt the following form when $T=0$  and $T \gg \Lambda_{QCD}$:
\begin{equation}\label{potential}
	V_{QCD}(a)=f^2_a m^2_a(T) \left[1- \cos{\left(\frac{a}{f_a}\right)} \right].
\end{equation}

It is of interest for the present work that the field theory of an axion field interacting with the gluonic fields by Eq.~(\ref{potential}) has a $Z_{N_{DW}}$ symmetry subgroup if $N_{DW} > 1$, with associated $N_{DW}$ degenerate vacua, of the Peccei-Quinn symmetry
that leaves the potential $V_{QCD}(a)$ invariant \cite{Peccei2008,Sikivie2008}, namely:
\begin{equation}
 \theta(x) \rightarrow \theta(x) + 2 \pi k, \qquad k=0,1,...,N_{DW}-1.
\end{equation}
This discrete symmetry is spontaneously broken when the axion field acquires a vacuum expectation value, which happens near to the temperature $T \sim \Lambda_{QCD}$,  leading to the formation of domain walls \cite{PhysRevD.85.105020}.
Generally, a domain wall is a two-dimensional topological defect separating two domains of a field and are indeed associated to the spontaneous breaking of a discrete symmetry  \cite{vilenkin1994cosmic,Sikivie2008}.

The $N_{DW}$ number is called "domain wall number" in the literature.
As treated in several references, e.g. Refs.~\cite{Sikivie2008,PhysRevD.85.105020,PhysRevD.32.1560,PhysRevD.30.712}, the creation of domain walls is very sensitive for the evolution of the axion field during the earliest stages of the Universe, overall for scenarios where $N_{DW}>1$ and the spontaneous breaking of the Peccei-Quinn symmetry happens after inflation.
The scenario described by the formerly cited references seems to lead to an early Universe dominated by axion domain walls, with an energy density much bigger than the critical one in the current Universe.

In the following (namely Section (\ref{dio})) we consider the possibility whether electromagnetic radiation could significantly contribute to the decay of domain walls and we enforce what is claimed in Ref.~\cite{PhysRevD.32.1560}, i.e. there cannot be any significant contribution.
In the following we describe a laboratory QCD axion field by assuming a coherent time-oscillating axion background, namely $\Theta(t)=\Theta_0 sin(\omega_a t)$. The latter is a very good approximation when the typical spatial dimension of the system $L$ is much smaller than the de Broglie wavelength of the axion field (typically of the order of $ 10^{-3} \; m_a^{-1}$ if we take $v \sim 10^{-3}$ and so $\omega_a \sim m_a$ cf.,   Millar et al. \cite{Millar_2017}).

\subsection{Axion-Modified Maxwell equations}
It is straightforward \cite{Millar_2017} to see that the interaction term in Eq.~(\ref{eq0}) can be written as an interaction Lagrangian of the form $ -J^{\nu} A_{\nu} $ by taking
\begin{equation}
	J^{\nu}=g_{a \gamma \gamma} \tilde{F}^{\mu \nu} \partial_{\mu}a =\frac{1}{2} g_{a \gamma \gamma} \partial_{\mu} a \; \epsilon^{\mu \nu \rho \sigma} \partial_{\rho} A_{\sigma}.
\end{equation}
The 4-current generated by axions is thus $J_{a}=(\rho_{a}, \vec{J}_{a}) $, where
\begin{equation}
	\begin{split}
	\rho_{a}&=g_{a \gamma \gamma} \vec{B} \cdot \nabla a,\\
	\vec{J}_{a}&= g_{a \gamma \gamma} \nabla a \wedge \vec{E}- g_{a \gamma \gamma} \dot{a} \vec{B}.
	\end{split}
\end{equation}
The continuity equation $  \dot{\rho}_{a}+\nabla \cdot \vec{J}_{a}=0 $
expresses the same as  $\partial_{\mu} j_a^{\mu}=0$.
Another straightforward way to see this is by observing from the structure of Eq.~(\ref{eq0}) that it is possible to redefine the tensor $H$ as:
\begin{equation}
	H^{a}_{\mu \nu}=H_{\mu \nu}-g_{a \gamma \gamma} a(x)\tilde{F}_{\mu \nu} ,
\end{equation}
in order to rewrite Eq.~(\ref{eq0}) with the interaction term $\frac{1}{4} g_{a \gamma \gamma} a \tilde{F}_{\mu \nu} F^{\mu \nu}$ included inside the 'free term' $-\frac{1}{4} F^{\mu \nu} H_{\mu \nu}$.
We then find that we can use different constitutive relations for $\vec{D}$ and $\vec{H}$ in Axion Electrodynamics:
\begin{subequations}
	\begin{align}\label{HD}
		\vec{D}_a=\varepsilon \vec{E}+\vec{P}_a,\\
		\vec{B}=\mu \vec{H}_a+\vec{M}_a,
	\end{align}
\end{subequations}
where we have defined the axion-induced polarization vector $\vec{P}_a=-g_{a \gamma \gamma} a \vec{B}$ and magnetization $\vec{M}_a=-\mu\,g_{a \gamma \gamma} a \vec{E}$,
which can be interpreted as polarization and magnetization vectors that are induced by the interaction with the axion and lead to effective "bound charge and current densities" $\rho_a=-\nabla \cdot \vec{P}_a $ and $\vec{J}_a=\nabla \wedge \vec{M}_a+\dot{P}_a$.
This means that the interaction between the axion and the electromagnetic field generates an effective polarization and magnetization, analogously to gravitational backgrounds. 

The Euler-Lagrange equations associated with the Lagrangian density (\ref{eq0}) are
\begin{subequations}
	\begin{align} \label{eq01}
		\ddot{a}-\nabla^2 a+V'(a)+g_{a \gamma \gamma} \vec{E} \cdot \vec{B} &=0,
	\end{align}
	\begin{align}\label{eq02}
		\nabla \cdot \vec{E}= \rho+g_{a \gamma \gamma} \vec{B} \cdot \nabla a,
	\end{align}
	\begin{align}\label{eq03}
		\nabla \wedge\vec{B}=\vec{J}+\dot{\vec{E}}-g_{a \gamma \gamma} \dot{a} \vec{B}+g_{a \gamma \gamma}  \nabla a \wedge \vec{E},
	\end{align}
	\begin{align}\label{eq04}
		\nabla \cdot \vec{B}=0,
	\end{align}
	\begin{align}\label{eq05}
		\nabla \wedge \vec{E}=-\dot{\vec{B}}.
	\end{align}
\end{subequations}
Equation (\ref{eq01}) is the Sine-Gordon equation for the axion field  interacting  with the electromagnetic field, while equations (\ref{eq02}) and (\ref{eq03}) are the modified Maxwell equations with sources and the equations (\ref{eq04}) and (\ref{eq05}) are the "constraint Maxwell equations" that are unchanged because they follow from the  Bianchi identities $\partial_{\mu} \tilde{F}^{\mu \nu}=0$.

The fields in (\ref{HD}) satisfy the following forms of Maxwell equations:
\begin{subequations}\label{Mxax}
	\begin{align}
		\nabla \cdot \vec{D}_a&= \rho,\\
		\nabla \wedge\vec{H}_a&=\vec{J}+\dot{\vec{D}}_a,\\
		\nabla \cdot \vec{B}&=0,\\
		\nabla \wedge \vec{E}&=-\dot{\vec{B}}.
	\end{align}
\end{subequations}
\subsection{Axion Electrodynamics in Condensed Matter}
As explained in Refs.\cite{nenno2020axion,PhysRevB.86.115133}, the electromagnetic properties of a Weyl semimetal can be described by a Chern-Simons theory with Lagrangian (\ref{eq0}), where the effective axion field is
\begin{equation}
	\Theta(x)=b_0 t-\vec{b} \cdot \vec{r}=b_{\mu} x^{\mu}.
\end{equation}
The quantities $b_{\mu} $ are of relevant physical meaning for the material, since they are related to the energy shift $b_0$ and the momentum shift $\vec{b}$ of the specific Weyl point of the material and whose Hamiltonian is
\begin{equation}
	h_W(\vec{k})=b_0+ v \vec{\sigma} \cdot (\vec{k}-\vec{b}).
\end{equation}
A Weyl semimetal constitutes an interesting phase of topological quantum matter with fascinating physical properties such as protected surface states and the electromagnetic response.

In the following, in order to consider an order of magnitude for $b_{\mu}$ in the case of time-dependent axion field and space-dependent one, we consider $b_0 \sim 10^{-1} \, \SI{}{eV}$ and $|\vec{b}| \sim 10 \, \SI{}{eV}$.
In the first case, the time derivative of the axion field is constant, so it is easier to deal with since e.g. the Green's function is dependent on $t-t'$, and can be regarded as an approximation of the second case when we consider times $0<t \ll \omega_a^{-1}$. The second case is a coherently oscillating  axion field, 
which can appropriately approximate a current laboratory axion field when the typical spatial dimension $L$ of the experimental system is much smaller than the de Broglie wavelength of the axion field (see Refs.~\cite{Millar_2017, RevModPhys.93.015004}).

\subsection{Electromagnetic 4-potential in Axion Electrodynamics}
As in usual electrodynamics, the differential  equations for the 4-potential $A_{\mu}(x)=(\phi,\vec{A})$ depend on  the choice of gauge.
We can in general write the electric and magnetic fields as
\begin{equation}
		\vec{E}=-\nabla \phi-\frac{\partial \vec{A}}{\partial t} \qquad\vec{B}=\nabla \wedge \vec{A}.
\end{equation}
Consequently, we have in the vacuum:
\begin{align}\label{equA}
	\nabla^2 \Phi+\partial_t \left(\nabla \cdot \vec{A}\right)&=-\rho_a,\\
	( \partial_t^2-\nabla^2) \vec{A}+ \nabla \left( \nabla \cdot \vec{A}+ \partial_t \Phi\right)&=\vec{J}_a.
\end{align}
In the following we will use mainly
the Coulomb gauge (radiation gauge if $\rho_a=0$)
	$\nabla \cdot \vec{A}=0$
and the temporal gauge
	$A_0=\Phi=0.$
In the case of temporal gauge we have:
\begin{equation}\label{temp}
	\Box \vec{A}+\nabla(\nabla \cdot \vec{A})=-\dot{\Theta} \; \nabla \wedge \vec{A}-\nabla \Theta \wedge \frac{\partial \vec{A}}{\partial t}.
\end{equation}

In the case of the radiation gauge with a time-dependent axion field we get the equations
	\begin{equation}\label{radiation}
		( \varepsilon \mu \partial_t^2-\nabla^2) \vec{A}= - \mu\dot{\Theta}\;  \nabla \wedge \vec{A}.
	\end{equation}
\section{Energy-momentum balance with fixed and dynamical axion background}\label{tensor}It is then easy to notice from the Lagrangian (\ref{eq0}) that the total energy-momentum balance reads (with $J^{\mu}=0$):
\begin{equation}\label{cons0}
	\partial_{\mu} T^{\mu \nu}=0,
\end{equation}
where we have that $T^{ \mu \nu}$ is simply the sum of the free Minkowski stress-energy tensor $T^{ M\mu \nu}=-\frac{1}{4} \eta^{\mu \nu} F_{\alpha \beta} H^{\alpha \beta}+F^{\mu}_{\rho} H^{\nu \rho}$ and the free axion one  $T^{\mu \nu}_{a}=\partial^{\mu} a \,\partial^{\nu} a- \eta^{\mu \nu} (\frac{1}{2} \partial_{\rho} a \,\partial^{\rho} a-\frac{1}{2} m_a^2 a^2)   $.
This simple result is a consequence of the fact that in this case the system formed by the electromagnetic field and the axion field is treated as a closed system (this is not rigorously exact since the potential $V(a)$ comes from the effective interaction with the gluonic fields). However, if we take the two fields alone such a relation as Eq.(\ref{cons0}) is not valid since they are open systems interacting between each other.
If we consider the electromagnetic field alone, one gets according to Ref.~\cite{doi:10.1142/S0217751X22501512}:
\begin{equation}
	\partial_{\nu}T^{M \nu}_{\mu}=-f^M_{\mu},
\end{equation}
where $f^M_{\mu}=(f^M_0,\vec{f}^M)$ whose spatial components are the components of Abraham's force density $
	\vec{f}^A= (\varepsilon \mu-1) \frac{\partial}{\partial t} \Big(\vec{E} \wedge \vec{H}\Big)-(\vec{E} \cdot \vec{B}) \nabla \Theta
$
and $f^M_0=-g_{a \gamma \gamma } \dot{a} \vec{E} \cdot \vec{B}$.

\section{Green's functions of Axion Electrodynamics}\label{vacgreen}
We first develop the Green's function approach in Axion Electrodynamics, using the temporal gauge.
We start from Eq.~(\ref{temp}) with $\varepsilon=\mu=1$ and define the kernel $ \vec{G}_{ij}(x,x')$ such that:
\begin{equation}\label{dyad}
	A_i(x)=\int d^4 x' \; G_{ij}(x,x') \, J^{j}(x').
\end{equation}
Due to causality, the variable $t'$ is only integrated over $t' \leq t$. From Eq.~(\ref{temp}) we find
\begin{equation}\label{greenaqua}
	\begin{split}
		\Box G_{ij}(x,x')+\partial_i(\partial_k G_{kj}(x,x'))-\dot{\Theta}(x) \epsilon_{ikl} \partial_k G_{lj}(x,x')-\epsilon_{ikl} \partial_k \Theta(x) \; \partial_t G_{lj}(x,x')=\\=\delta_{ij} \delta^{(4)}(x^{\mu}-x'^{\mu}).
	\end{split}
\end{equation}

The $G$ kernel (\ref{dyad}) is equal to $ iD_{i j}(x^{\mu} ,y^{\mu})$ where $D$ is the retarded Green's function of the vector potential $\vec{A}$, defined as
\begin{equation}\label{GR}
	i D^{R}_{i j}(x^{\mu}, y^{\mu})=\begin{cases}
		\langle{A_{i}(x) A_{j}(y)-A_{j}(x) A_{i}(y)}\rangle \; \;  \; \textit{if} \;  \;  \;  \;x^{0} -y^{0} > 0,\\ 0 \;  \;  \;  \; \rm{otherwise},
	\end{cases}
\end{equation}
can be derived by the Schwinger-Dyson equations \cite{schwartz2014quantum} for Axion Electrodynamics which is indeed Eq.~(\ref{greenaqua}) with the temporal gauge. As shown in Refs.~\cite{landau2013electrodynamics,https://doi.org/10.48550/arxiv.hep-th/9901011} and \cite{birkeland2007feigel}, the calculation of the retarded Green's function  (\ref{GR})  is useful in order to get the significant two-point physical averages between electric and magnetic fields.
One can easily find that with this choice of gauge (valid also in radiation gauge):
\begin{subequations}\label{two}
\begin{align}
	\langle{E_{i}(x) E_{j}(x') }\rangle &=\partial_t \partial_{t'}\langle{A_{i}(x) A_{j}(x') }\rangle, \\
	\langle{B_{i}(x) B_{j}(x') }\rangle &=\curl_{il}  \curl'_{jm} \langle{A_{l}(x) A_{m}(x') }\rangle,  \\
		\langle{B_{i}(x) E_{j}(x') }\rangle &=-\curl_{il} \partial_{t'} \langle{A_{l}(x) A_{j}(x') }\rangle,
\end{align}
\end{subequations}
and similarly for other two-point functions. We also use in the following Fourier transforms (FT) of the two-point functions, e.g.
\begin{equation}
\langle{A_{\alpha}(x) A_{\beta}(y) }\rangle_{\omega} =\int_{-\infty}^{+\infty} dt \, e^{i \omega t} \langle{A_{\alpha}(t,\vec{x}) A_{\beta}(0,\vec{y}) }\rangle,
\end{equation}
and similarly for FTs, from which the relations (\ref{two}) can be used to find the corrispondent FTs for two-point functions involving electric and magnetic fields.\\
For example, if the Green's function is a function of the variable $t-t',x-x' \,\, \text{and} \,\, y-y'$ we can define the
reduced Green’s function $g_{ij}(z,z',\omega,\tilde{k})$:
\begin{equation}
	G_{i j}(x,x')=\int_{-\infty}^{\infty} \frac{d \omega}{2 \pi} \int_{-\infty}^{\infty} \frac{d^2 \tilde{k}}{(2 \pi)^2} e^{-i \omega (t-t')} e^{i \tilde{k} \cdot (\vec{r}-\vec{r}')} g_{ij}(z,z',\omega, \tilde{k}),
\end{equation}
where we use the notation $\tilde{k}=(k_x,k_y)$. We also adopt the useful quantity $\kappa^2=\omega^2-k_x^2-k_y^2.$
In the following we are more focused on the Green's function in the vacuum for calculating Casimir forces for our systems of interest. From this the following stress tensor component, important in a Casimir context, can be
found straightforwardly ($\varepsilon=\mu=1$):
\begin{equation}\label{Tzz}
	\begin{split}
		\langle{T^{M}_{zz}}\rangle_{\omega,\tilde{k}}=\frac{1}{2i} \left[-\kappa^2 g_{zz}+(\omega^2-k_y^2) g_{xx}+(\omega^2-k_x^2) g_{yy}+ik_{y} \left(\partial_z g_{yz}-\partial_z' g_{zy}\right) \right.\\ \left.  +ik_{x} \left(\partial_z g_{xz}-\partial_z' g_{zx}\right)
		+k_x k_y (g_{xy}+g_{yx}) + \partial_z \partial_z' (g_{xx}+g_{yy})         \right].
	\end{split}
\end{equation}
For our purpose of evaluating zero-point energies we are also using the retarded Green's function with $\mu \neq 1$ in order to find the vacuum one by using the fluctuation-dissipation theorem \cite{landau1987statistical}:
\begin{equation}
	\langle{A_{\alpha}(x) A_{\beta}(y) }\rangle_{\omega}=\frac{i}{2} \left[ D^{R}_{\alpha \beta}(\omega,\vec{x}, \vec{y})-D^{R*}_{ \beta \alpha}(\omega, \vec{y},\vec{x})  \right]
\end{equation}
and then make the limit $\mu \rightarrow 1+i0$.
Notice that in the case of temporal gauge, the Green's function is a $3\times 3$ tensor and not a $4\times 4$ one.
If the axion-generated charge density $\rho_a=0$, that is the case when the axion field is only a function of time $t$, it is convenient to adopt the radiation gauge and it is what we adopt in the following section. We adopt the temporal gauge for space-dependent axion fields from Section (\ref{spazio}) on.
\section{Purely time-dependent axion field}\label{time-dependent}
Here we observe that the only difference between the Green's function in temporal gauge and radiation gauge (with an axion field depending only on time) relies on the resulting equation for Green's function $G_{kj}(x,y)$ that becomes from Eq.~(\ref{radiation}), if we assume a homogenous medium with electrical permittivity $\epsilon=1$ and magnetic permeability $\mu $:
\begin{equation}\label{Green1time}
	\left[   \delta_{ik} \Box(\mu) -\mu \dot{\Theta}(t) \epsilon_{lik} \nabla_l \right] G_{kj}(x,y)=\mu \delta_{ij}  \delta^{(4)}(x^{\mu} -y^{\mu}),
\end{equation}
where the operator $\Box(\mu)= \mu \partial_t^2- \nabla^2$.
Since the time derivative of the axion field can be time-dependent, the solution of Eq.(\ref{Green1time}) is generally dependent on $ \vec{r}-\vec{r'}$ and $t-t'$, but explicitly also on $t$.
It is not trivial to find solutions with a generic time dependence of the axion field. In the following we work out two specific cases of interest we mentioned in Section \ref{intro}, i.e. $\Theta=\Theta_0 t$ and $\Theta=\Theta_0 \sin(\omega_a t)$.
\subsection{Constant axion time derivative $\dot{\Theta}$}\label{constdot}
If we consider $\dot{\Theta}$ as constant in time and uniform in space we can take, similarly to \cite{PhysRevD.102.123011}, the FT of the Green's function and get from Eq.~(\ref{Green1time}) the following equation:
\begin{equation}\label{Greenconst}
	\left[ \delta_{ik} (-\mu \omega^2+ k^2)+i \mu \dot{\Theta} \epsilon_{lik} k_l \right]  \tilde{G}_{ k j}(\vec{k}, \omega)=\mu \delta_{ij}.
\end{equation}
The calculations in this case are of conceptual  interest and can be applied in a similar manner in  a physically more interesting case, such as the oscillating axion field. We will come back to this topic  in the next section.

In order to get the Green's function with vacuum surroundings we can put a magnetic permeability $\mu=1+i \epsilon(\omega)$, where $\epsilon(\omega)$ is an infinitesimal of the first order, and we evaluate the limit for $\epsilon \rightarrow 0$ in the proper order, as analogously done in Ref.~\cite{landau1987statistical} with an electrical permeability $\varepsilon(\omega)=1+i \epsilon(\omega)$ to get the quantum mechanical result for the Green's function.
A magnetic permeability of the form $\mu=\mu_r+ i \mu'$ can  be found, e.g. for real media where magnetic viscosity is present (see e.g. Ref.~\cite{WanjunKu1997}):
\begin{equation}
	\mu=\Re{(\mu)}+i \epsilon \frac{\omega}{\omega_0},
\end{equation}
which is analogous to the case of electric permittivity, where we have the imaginary part $\Im{(\mu)}>0$ when $\omega>0$. This also justifies the presence of $\sgn(\omega)$.

From Eq.~(\ref{Greenconst}) we get, similarly to Ref.~\cite{PhysRevD.102.123011}, the retarded Green's function FT as the inverse operator of the expression in square brackets on the left side in the mentioned limit of $\mu=1+i \epsilon \sgn(\omega) \rightarrow 1+i0 \sgn(\omega)$:
\begin{equation}\label{solutotutto}
	\tilde{G}_{jk}(\vec{k}, \omega)= (\omega^2-|\vec{k}|^2)\tilde{A}(\omega, \vec{k},\vec{\beta}) \delta_{jk}+i\tilde{A}(\omega, \vec{k},\vec{\beta}) \epsilon_{jkl}\beta_l+\frac{1}{\omega^2-|\vec{k}|^2+i 0 \sgn(\omega) }\tilde{A}(\omega, \vec{k},\vec{\beta}) \beta_j \beta_k,
\end{equation}
where $\vec{\beta}$ is a vector with components $	\beta_j=\dot{\Theta} k_j $, whose squared module is $\beta^2=\dot{\Theta}^2 |\vec{k}|^2$ and
\begin{equation}
		\tilde{A}(\omega, \vec{k},\vec{\beta})=\frac{1}{(\omega^2-|\vec{k}|^2)^2-\beta^2+i 0 \sgn(\omega) } .
\end{equation}
The ordinary electrodynamics limits can be obtained by setting $\dot{\Theta}=0$.
Those Fourier transforms we got can be helpful for two applications we develop in the following, namely for evaluating the spectral energy density $\rho_{em}(\omega)$ of the electromagnetic field at temperature $T$, the Casimir force between two parallel conducting plates in such a fixed axionic background and a discussion on dispersion relations.
\subsubsection{Spectral energy density $\rho_{em}(\omega)$ of electromagnetic field}
In order to get the zero-point spectral energy density we calculate the FT $\tilde{G}(\vec{r},\omega) $ of the Green's function (\ref{solutotutto}).
When calculating energy density we need the correlation functions in the limit $\vec{r} \rightarrow 0$.
The usual spectral energy density of the electromagnetic field is
\begin{equation}
	\rho(\omega,\vec{r}) d \omega= \frac{1}{2} \left[2 \langle{\vec{E}^2(\vec{r})}\rangle_{\omega,T}+2\langle{\vec{B}^2(\vec{r})}\rangle_{\omega,T} \right]  \frac{d \omega}{2 \pi},
\end{equation}
where the factors 2 inside the brackets are inserted because we follow the  definition of spectral densities given in  Ref.~\cite{landau1987statistical}:  average spectral values  are defined  as integrals in $\omega$ from $-\infty$ to $+\infty$.
We also exploit the relation:
\begin{equation}\label{distro}
	\frac{1}{x \pm i 0}=\mathcal{P}{\frac{1}{x}} \mp i \pi \delta(x).
\end{equation}
We expand the  spectral energy density in a perturbative series:
\begin{equation}
	\rho(\omega,\vec{r})=\rho^{(0)}(\omega,\vec{r})+\rho^{(1)}(\omega,\vec{r})+\rho^{(2)}(\omega,\vec{r})+...
\end{equation}
according to the perturbative factor, that we can take here to be $\dot{\Theta}$. We can develop a perturbative expansion for the Green's function:
\begin{equation}
	\tilde{G}_{jk}(\vec{k}, \omega) \sim \tilde{G}^{(0)}_{jk}(\vec{k}, \omega)+\tilde{G}^{(1)}_{jk}(\vec{k}, \omega)+\tilde{G}^{(2)}_{jk}(\vec{k}, \omega),
\end{equation}
where
\begin{subequations}\label{consthetaG}
	\begin{align}
		\tilde{G}^{(0)}_{jk}(\vec{k}, \omega)&=\frac{1}{\omega^2-k^2+i0 \sgn{\omega}}, \\
		\tilde{G}^{(1)}_{jk}(\vec{k}, \omega)&=\frac{i}{(\omega^2-k^2)^2+i0 \sgn{\omega}} \epsilon_{jkl}\beta_l=\frac{i}{(\omega^2-k^2)^2+i0 \sgn{\omega}} \epsilon_{jkl} (\dot{\Theta} k_l),\\
		\tilde{G}^{(2)}_{jk}(\vec{k}, \omega)&=\frac{\beta^2}{(\omega^2-k^2)^3+i0 \sgn{\omega}}\left[ \delta_{jk}+\frac{\epsilon_{jkl}\beta_l}{\omega^2-k^2} \right]+\frac{1}{(\omega^2-k^2)^3+i0 \sgn{\omega}}\dot{\Theta}^2 k_j k_k.
	\end{align}
\end{subequations}

As shown in Ref.~\cite{landau1987statistical} it is easy to obtain (by integrating over the domain of $\vec{k}$) that the spectral density associated with the zero order term $\rho^{(0)}(\omega)$ is
\begin{equation}
	\rho^{(0)}(\omega) d \omega=  d \omega \; \omega^2 \left[\frac{1}{2} \omega + \frac{\omega}{e^{\frac{\omega}{T}}-1} \right].
\end{equation}
This is the familiar result from QED: the zero-point energy of electromagnetic field at a given temperature $T$ is the sum of the zero-point energy at the zero Kelvin temperature and the black-body radiation energy at the given temperature $T$.
The first order term $\rho^{(1)}(\omega)$ is trivially zero, while we have a contribution of  order $\mathcal{O}(g^2_{\alpha \gamma \gamma}) $:
\begin{equation}\label{dothecon}
	\rho^{(2)}(\omega)\; d \omega=  d \omega \; \omega^2 \left[\frac{1}{2} \frac{ \dot{\Theta}^2}{8\omega} + \frac{1}{e^{\frac{\omega}{T}}-1}\frac{ \dot{\Theta}^2}{8\omega} \right].
\end{equation}
We notice in Eq.~(\ref{dothecon}) the frequency dependence of the  spectral energy density, from which we can find the  spectral emissivity using Planck's law. The first term on the right hand side  is proportional to the frequency. As $\dot{\Theta}$ can be regarded small for physical purposes,  this  frequency dependence suggests that this can be physically interpreted as a blue noise.

The second term is similar to the first one  as regards the frequency dependence,  but it has a  Bose-Einstein temperature-dependent weight. For $T \gg \omega$ it is term going as $\simeq T $, so it can be regarded as a flat noise, and surely is bigger than the zero-temperature term. For $T \ll \omega$ the Bose-Einstein weight would be approximately $e^{-\frac{\omega}{T}}$, so it gives an exponential noise.
\subsubsection{Casimir force between parallel plates with time-increasing axion field}
Here we evaluate the axion modifications to the usual Casimir force between two parallel perfectly conducting plates (see Refs.~\cite{https://doi.org/10.48550/arxiv.hep-th/9901011,schwartz2014quantum} for a treatment without an axion background).
Our result is comparable to the one obtained by Ref.~\cite{PhysRevD.81.025015}, while improving their result.

We can evaluate the zero-point energy $u_{em}(L,\dot{\Theta})$ per unit transverse area by evaluating the following expression:
\begin{equation}\label{uem}
	u_{em}(L,\dot{\Theta})=\frac{1}{2} \sum_{\pm} \sum_n \int \frac{d^2 k}{(2 \pi)^2} \sqrt{|\vec{k}|^2+ \frac{n^2 \pi^2}{L^2} \pm \dot{\Theta} \sqrt{|\vec{k}|^2+ \frac{n^2 \pi^2}{L^2}  }}.
\end{equation}
This could be expected by a Casimir approach, namely by physical qualitative means given by the dispersion relations  .
The expression (\ref{uem}) can be demonstrated overall by a Green's function approach. We assume the $z$-axis to be the direction normal to the plates.
We need to solve the Green's function equations with the proper boundary conditions. Since we assume to have two perfect conducting plates we have boundary conditions for electric field and magnetic field at $z=0,L$:
\begin{equation}
B_z=0 \qquad \vec{E}_{//}=0,
\end{equation}

where $\vec{E}_{//}$ is the parallel component of the electric field, from which we get the correspondent ones for the Green's function:
\begin{subequations}\label{BCs}
	\begin{align}
g_{ij}(x-x',y-y',z,z')|_{z=0,z=L}&=0	\quad i \neq z, j \neq z,\\
\partial_z g_{zz}(x-x',y-y',z,z')|_{z=0,z=L}&=0.
	\end{align}
\end{subequations}
These conditions, along with translational invariance along x and y directions,  lets us write the Green's function in the following Fourier form for $0<z<L$ and $0<z'<L$:
\begin{subequations}\label{fouseries}
	\begin{align}
	G_{ij}(x,x')&=\int_{-\infty}^{+\infty} \frac{d \omega}{2 \pi}\int_{-\infty}^{\infty} \frac{d^2 \tilde{k}}{(2 \pi)^2} e^{i \tilde{k} \cdot (\tilde{r}-\tilde{r}')} \frac{2}{L}\sum_{n=0}^{\infty}	\sin{\left(\frac{n \pi}{L}z\right)} \sin{\left(\frac{n \pi}{L}z'\right)} \tilde{g}_{ij}(\tilde{k},n) \quad i \neq z, j \neq z,\\
	G_{zz}(x,x')&=\int_{-\infty}^{+\infty} \frac{d \omega}{2 \pi}\int_{-\infty}^{\infty} \frac{d^2 \tilde{k}}{(2 \pi)^2} e^{i \tilde{k} \cdot (\tilde{r}-\tilde{r}')} \frac{2}{L}\sum_{n=0}^{\infty}	\iota(n)\cos{\left(\frac{n \pi}{L}z\right)} \cos{\left(\frac{n \pi}{L}z'\right)} \tilde{g}_{zz}(\tilde{k},n),
	\end{align}
\end{subequations}
where $\tilde{k}=(k_x,k_y)$, $\tilde{r}=(x,y)$ and $\iota(n)=1$ for $n>0$, while $\iota(n=0)=1/2$.\\
The reduced function $\tilde{G}(\tilde{k},n)$ consequently satisfies the equations (\ref{Greenconst}), with solution Eq.~(\ref{solutotutto}) ,where $k_z= \frac{n \pi}{L}$.
We show now two ways of demonstrating Eq.~(\ref{uem}), then we calculate the Casimir force.
\paragraph{Energy density per transversal area from $\Im{(\mu)} \rightarrow 0$}

Here we adopt the form (\ref{solutotutto}) for the Green's function.
If we exploit the relation (\ref{distro})
and use the Green's function (\ref{solutotutto}) we get
\begin{equation}
	\frac{1}{2}\langle{\vec{E}^2+\vec{B}^2}\rangle_{\omega,k_x,k_y,n} =\pi \frac{\omega}{2} \sum_{\pm} \left[\delta\left(\omega - \sqrt{|k|^2 \pm \dot{\Theta} |k|}\right)- \delta \left(\omega+ \sqrt{|k|^2 \pm \dot{\Theta} |k|}\right)      \right],
\end{equation}
where $ |k|^2=k_x^2+k_y^2+ \left(\frac{n \pi}{L}    \right)^2 $.
We then exploit the properties of the Dirac delta distribution by integrating in the $\frac{\omega}{2 \pi}$ domain from $-\infty$ to  $+\infty$ and we then get
\begin{equation}
		\frac{1}{2}\langle{\vec{E}^2+\vec{B}^2}\rangle_{k_x,k_y,n}=\frac{1}{2} \sum_{\pm} \sqrt{k_x^2+k_y^2+ \frac{n^2 \pi^2}{L^2} \pm \dot{\Theta} \sqrt{k_x^2+k_y^2+ \frac{n^2 \pi^2}{L^2}}}.
\end{equation}
We also mention that $	\langle{\vec{A} \cdot \vec{B}}\rangle_{k_x,k_y,n}$ and $	\langle{\vec{E} \cdot \vec{B}}\rangle_{k_x,k_y,n}$  are trivially equal to zero. Consequently, we have $	\langle{T^{00}_{em}}\rangle_{k_x,k_y,n}=	\frac{1}{2}\langle{\vec{E}^2+\vec{B}^2}\rangle_{k_x,k_y,n}$, from which it is easy to get the energy per transverse area unit to be equal to the expression (\ref{uem}).

\paragraph{Casimir force by the reduced Green’s function in the case $\mu=1$}
In general, the Casimir force is the physical force between dielectric or metallic surfaces (and in different configurations), caused by quantum fluctuations of the electromagnetic field. 
From a general perspective, the Casimir effect can be treated by different methods. 

It can be derived from statistical equilibrium thermodynamics, as an  application of the fluctuation-dissipation theorem, which relates the two-point functions of the  electromagnetic field to the imaginary part of retarded Green function, as discussed in Landau-Lifshitz book on statistical physics \cite{landau1987statistical} and mentioned before, as well as in several articles, for instance Ref.\cite{Brevik_Shapiro_Silveirinha_2022}. Alternatively , the effect can be seen upon as a manifestation of quantum field theory; cf. Milton's book \cite{https://doi.org/10.48550/arxiv.hep-th/9901011}.

In both cases, the essential ingredient required in practical calculations is the difference between the normal Maxwell stress components  $T_{zz}$ on the two sides of a boundary.
We explicitly calculate the expression (\ref{Tzz})  with the Green's function solving the equation (\ref{Greenconst}) in the form (\ref{fouseries}) when exactly $\mu=1$. It is easy to get:
\begin{equation}\label{tzo1}
	\langle{T_{zz}}\rangle_{\omega, k_x,k_y}|_{z=z'=0,L}= - 2 i \sum_{n=1}^{+\infty }\frac{n^2 \pi^2}{L^3} \frac{\kappa^2-\frac{n^2 \pi^2}{L^2}}{(\kappa^2-\frac{n^2 \pi^2}{L^2})^2-\dot{\Theta}^2 (k_x^2+k_y^2+\frac{n^2 \pi^2}{L^2})},
\end{equation}
We remind that $\kappa^2=\omega^2-k_x^2-k_y^2$.
This series can be treated by noticing that the $n$-term of the series can be written as:
\begin{equation}\label{tzo2}
\frac{\frac{n^2 \pi^2}{L^2}(\kappa^2-\frac{n^2 \pi^2}{L^2})}{(\kappa^2-\frac{n^2 \pi^2}{L^2})^2-\dot{\Theta}^2 (k_x^2+k_y^2+\frac{n^2 \pi^2}{L^2})}=\frac{1}{2} \sum_{\pm} \frac{\frac{n^2 \pi^2}{L^2}}{\kappa^2-\frac{n^2 \pi^2}{L^2} \pm \dot{\Theta} \sqrt{k_x^2+k_y^2+\frac{n^2 \pi^2}{L^2}}},	
\end{equation}
and one can perform a standard complex frequency rotation $\omega \rightarrow \zeta=i \omega$ , similarly to what done in Ref.~\cite{https://doi.org/10.48550/arxiv.hep-th/9901011}, and integrate over $\zeta/{2 \pi}$, obtaining:
\begin{equation}\label{tzo3}
		\langle{T_{zz}}\rangle_{k_x,k_y}|_{z=z'=0,L}= \sum_{\pm} \sum_{n=1}^{+\infty }\frac{n^2 \pi^2}{L^3} \frac{1}{\sqrt{k_x^2+k_y^2+\frac{n^2 \pi^2}{L^2}\pm\dot{\Theta} \sqrt{k_x^2+k_y^2+\frac{n^2 \pi^2}{L^2}}}}.
\end{equation}
It is easy to see from Eqs.~(\ref{tzo1},(\ref{tzo2}) and (\ref{tzo3}) that, also up to contact terms, we are dealing with divergent expressions. This is because the physical quantity we observe (Casimir force) is the discontinuity of $T_{zz}$, not the single values on the two sides of the interface(at $z=0^+$ as in this case).
We then need to consider $T_{zz}$ at $z=0^{-}$.

In order to do that, one needs to solve Eq.~(\ref{Green1time}) with $z<0$ and $z'<0$ (the case with $z>L$ gives the same results from symmetry) with the boundary conditions (\ref{BCs}) at z=0 and $g \sim e^{-ikz}$ for $z \rightarrow -\infty$. This can be done in a similar fashion to what done in Stakgold's book \cite{stakgold2011green}, by noticing that, accordingly to the boundary condition at z=0, the sine (or cosine) Fourier transform satisfies Eq.(\ref{Greenconst}).

We can then solve the equations, as done in Ref.~\cite{stakgold2011green}, and get the following $T_{zz}$ expression:
\begin{equation}\label{tzo}
	\langle{T_{zz}}\rangle_{\omega, k_x,k_y}|_{z=z'=0^-}= - 2 i \int_{0}^{+\infty} k_z^2 \frac{\kappa^2-k_z^2}{(\kappa^2-k_z^2)^2-\dot{\Theta}^2 (k_x^2+k_y^2+k_z^2)} dk_z.
\end{equation}
We manipulate this expression in a similar fashion to the one at $0^+$,integrate over $\zeta$ and  omit contact terms. We then get the Casimir force as the discontinuity of $T_{zz}$:
\begin{equation}
f(L,\dot{\Theta})=   \int_{-\infty}^{\infty} \frac{dk_x}{2 \pi} \int_{-\infty}^{\infty} \frac{dk_y}{2 \pi}  \frac{1}{L}\sum_{n=1}^{+\infty}  \sum_{\pm} \left[  \frac{\frac{n^2 \pi^2}{L^2} \pm \dot{\Theta} \frac{n^2 \pi^2}{ L^2 \sqrt{k_x^2+k_y^2+\frac{n^2 \pi^2}{L^2}}} }{\sqrt{|\vec{k}|^2+n^2 \pi^2/L^2\pm \dot{\Theta} \sqrt{|\vec{k}|^2+ \frac{n^2 \pi^2}{L^2}  }}}\right].
\end{equation}
It is easy to notice that the following formal relation holds:
\begin{equation}
	f(L,\dot{\Theta})=-\frac{\partial u_{em}(L,\dot{\Theta})}{\partial L},
	\end{equation}
confirming our results.

\paragraph{Calculation of the Casimir force}
In order to evaluate the Casimir force deriving from the expression (\ref{uem}) it is convenient to adopt the zeta function regularization method and extending it to a dimension $d \neq 2$  (see e.g. Milton \cite{https://doi.org/10.48550/arxiv.hep-th/9901011} and Brevik \cite{universe7050133}).
 We employ the Schwinger proper-time representation for the square root:
\begin{equation}\label{ue}
u_{em}=\frac{1}{2} \sum_{\pm} \sum_n \int \frac{d^d k}{(2 \pi)^d} \int_0^{+\infty} \frac{dt}{t} t^{-\frac{1}{2}} e^{-t\left(k^2+n^2 \pi^2/L^2\pm \dot{\Theta} \sqrt{|\vec{k}|^2+ \frac{n^2 \pi^2}{L^2}  }\right)} \frac{1}{\Gamma(-\frac{1}{2})},
\end{equation}
It is not easy to evaluate it exactly, so we will calculate  it perturbatively up to second order in $\dot{\Theta}$.
It is possible to write the Taylor series of the exponential in Eq.~(\ref{ue}) up to the second order in $\dot{\Theta}$ and sum up in the two polarizations explicitly:
\begin{equation}
	u_{em} \simeq \sum_n \int \frac{d^d k}{(2 \pi)^d} \int_0^{+\infty} \frac{dt}{t} t^{-\frac{1}{2}} e^{-t\left(k^2+n^2 \pi^2/L^2 \right)} \left[2+ t^2  \dot{\Theta}^2 \left(k^2+ \frac{n^2 \pi^2}{L^2}\right)\right] \frac{1}{\Gamma(-\frac{1}{2})}.
\end{equation}
As it is usually done in this kind of calculations, we carry out the Gaussian integration over k,  use the Euler representation of Gamma function, carry out the sum over $n$ via the definition of Riemann zeta function and we get:
\begin{equation}
	u_{em} \simeq u_0(L,d) +u_{ax}(L,d,\dot{\Theta}),
\end{equation}
where
\begin{subequations}
\begin{align}
	u_0(L,d)&=-\frac{1}{2 \sqrt{\pi}} \frac{1}{(4 \pi)^{d/2}} \left( \frac{\pi}{L}\right)^{d+1}  \Gamma \left(-\frac{d+1}{2}\right) \zeta(-d-1) ,\\
u_{a}(L,d,\dot{\Theta})&=-\frac{\dot{\Theta}^2}{2 \sqrt{\pi}} \frac{1}{(4 \pi)^{d/2}}  \left(\frac{\pi}{L}\right)^{d-1}  \left[ \Gamma \left(\frac{3-d}{2} \right) \zeta(-d-1)+\frac{d}{2} \, \Gamma \left(\frac{1-d}{2} \right) \zeta(-d+1)  \right].
\end{align}
\end{subequations}
The first term $u_0$ is the familiar result of Quantum Electrodynamics, while $u_{a}$ is the additional second-order axion contribution. For $d=2$ they are equal to \begin{subequations}\label{uao}
\begin{align}
u_0(L,d=2)=-\frac{\pi^2}{720} \frac{1}{L^3} \, \, \, ,	\qquad 
u_{a}(L,\dot{\Theta}, d=2)=-\dot{\Theta}^2 \frac{7}{320} \frac{1}{L} \, \, \,,
\end{align}
\end{subequations}
from which:
\begin{subequations}
	\begin{align}
		f_0(L)=-\frac{\pi^2}{240} \frac{1}{L^4}  \, \, \, ,\qquad
		f_a(L,\dot{\Theta})=-\frac{7 \dot{\Theta}^2}{320} \frac{1}{L^2} \, \, \,.
	\end{align}
\end{subequations}
It is worth noticing that we get  a second-order axion-induced modification of the  Casimir energy per transverse area going as $\sim \frac{1}{L}$, similarly to what is obtained in Ref.~\cite{PhysRevD.81.025015}, although our result differs by a multiplicative factor of $0.63$ (our result is lower). There are some points to notice about the possibilities of physical application of these results.

The approximation $\dot{\Theta} \sim $ constant for the fundamental physics axion field could be applicable to approximate a coherently oscillating axion field for very small times $t \ll \omega_a^{-1}$, meaning that the  theory is a reasonable approximation for frequencies $\omega \gg \omega_a $. It could then be used  when $\frac{1}{L} \gg \omega_a $.
This means that the treatment of the Casimir force between two conducting plates with such an axion background    is relevant for experimental purposes because the assumption for the perturbative approach is    $L \ll (\Theta_0 \, \omega_a)^{-1}  $, which is a stricter requirement than the long wavelength approximation, since for an axion being the main component of Dark Matter we need $\Theta_0 \sim 10^{-19}$ \cite{universe7050133, brevik2022electric} at cosmological scales.

If we take then the case of High Energy Physics axion at cosmological scales, $\dot{\Theta} \sim \Theta_0 \omega_a$, and a comparison between the terms in Eq.~(\ref{uao})  tells us that $u_a$ is comparable with $u_0$ when $L_a \sim (\Theta_0 \, \omega_a)^{-1} \sim 10^{14} \left( \frac{m_a}{10^{-2} \SI{}{eV}}\right) $ meters, so $u_a$ would be very small for usual Casimir experimental devices, if there are no significant local deviations in the axion field.
However, this axionic term could be significant in the case of topological insulators. Namely, if we take $b_0 \sim 10^{-1} \, \SI{}{eV}$, this corresponds to $\dot{\Theta} \sim 10^{-1} \,\SI{}{eV}$, so now $L_a \sim \SI{2}{\mu m}$.

\subsubsection{Dispersion relations}
Here we briefly discuss the dispersion relations for the case of a constant $\dot{\Theta}$. From the Green's function   it is easy to find the 4 poles to which they are associated with the dispersion relations:
\begin{equation}\label{equodisperdo}
	\omega_{\pm}=\sqrt{|\vec{k}|^2 \pm \dot{\Theta} |\vec{k}| } \qquad
	\omega_0=|\vec{k}|.
\end{equation}
This last observation can be understood in the following way: one firstly consider the Fourier equations of vector potential for a plane wave propagating along the direction of the $z$-axis:
	\begin{align}
		\begin{split}
			(-\omega^2+k_z^2) \mathcal{A}_x &=\dot{\Theta} k_z  \mathcal{A}_y,\\
			(-\omega^2+k_z^2) \mathcal{A}_y &=-\dot{\Theta} k_z  \mathcal{A}_x,	\\
			(-\omega^2+k_z^2) \mathcal{A}_z &=0.
		\end{split}
	\end{align}
This assumption on the propagation direction does not lose generality, since the treated system in the vacuum is isotropic and it is possible to find an inertial reference frame, where the electromagnetic wave propagates along the $z$-axis.
It is then easy to demonstrate the dispersion relation $\omega_{\pm}$ of (\ref{equodisperdo}) for real transverse photons. Indeed, if we define the fields as $\mathcal{A}_{\pm}=\mathcal{A}_x \pm i \mathcal{A}_y$, one gets the dispersion relations $\omega_{\pm}$ with $k_x=k_y=0$ . One then gets the expression for $\omega_{\pm}$ in Eq.~(\ref{equodisperdo}) by using the relativistic invariance of the magnitude of the wave 4-vector. The physical meaning of this is interesting: a left-circular polarized wave has a different frequency than a right-circular one with the same $\vec{k}$. This is due to the optical activity of the vacuum in Axion Electrodynamics as we shall discuss in Section~\ref{optical}.

We also notice that a solution for the equation of $\mathcal{A}_z$, the one with the dispersion relation $\omega=|\vec{k}|$ and corresponding to longitudinal photons, is simply zero, so that the dispersion relation corresponds to virtual photons.

\subsection{High frequency approximation for the Green's function}
Before working out the case of an oscillating axion field (by perturbative methods), we can get some understanding of the behavior by considering the High frequency approximation in that case. Start from  Eq.~(\ref{greenaqua}), substitute the form $\Theta(t)=\Theta_0 \sin{(\omega_a t)} $, and calculate the FT:
\begin{equation}\label{Green2time}
	(-\omega^2+|\vec{k}|^2) \tilde{G}_{ij}(\omega,\vec{k}) -i\epsilon_{lik}  k_l \frac{\Theta_0 \omega_a}{2} \left[ \tilde{G}_{ij}(\omega+\omega_a,\vec{k})+\tilde{G}_{ij}(\omega-\omega_a,\vec{k})  \right] =\delta_{ij}.
\end{equation}
We have here used the modulation property of the Fourier Transform in time.
The High frequency approximation consists on assuming $\omega \gg \omega_a$.
In this way, up to first order in $\omega_a$,  the Green's function satisfies the following equation,
\begin{equation}\label{Green2timeHF}
	(-\omega^2+|\vec{k}|^2) \tilde{G}_{ij}(\omega,\vec{k}) -i\epsilon_{lik}  k_l \Theta_0 \omega_a \tilde{G}_{kj}(\omega,\vec{k})   =\delta_{ij}.
\end{equation}
It has the same form as the preceding equation and has accordingly the  same solutions, with just the substitution $\dot{\Theta} \rightarrow \Theta_0 \omega_a$.
The interpretation of Eq.~(\ref{dothecon}) terms are still valid in this case.
\subsection{Case with $\Theta(t)=\Theta_0 \, sin(\omega_a t)$: Spectral energy density and production of real photons}
Now we treat a case of physical interest as mentioned before.
In order to treat this case properly  we need to work out  Eq.~(\ref{Green1time}) in more detail.\\
We assume that the wave vector is directed along a fixed direction, without losing generality. Actually, the axion field is taken to be dependent on time but spatially homogeneous, so it is isotropic.
We  make a FT in space and put  $\vec{k}=|\vec{k}| \hat{e}_z$,
\begin{equation}
	\begin{cases}
		(\partial_t^2+|\vec{k}|^2) \tilde{G}_{xx}(\vec{k},t,t')+i \dot{\Theta} k_z \tilde{G}_{zx}(\vec{k},t,t')=\delta(t-t'),\\
		(\partial_t^2+|\vec{k}|^2) \tilde{G}_{yy}(\vec{k},t,t')+ i \dot{\Theta} k_z \tilde{G}_{xy}(\vec{k},t,t')=\delta(t-t'),\\
		(\partial_t^2+|\vec{k}|^2) \tilde{G}_{zz}(\vec{k},t,t')=\delta(t-t'),\\
		(\partial_t^2+|\vec{k}|^2) \tilde{G}_{xy}+i\dot{\Theta}  k_z \tilde{G}_{yy}=0,\\
		(\partial_t^2+|\vec{k}|^2) \tilde{G}_{yx}-i\dot{\Theta} k_z  \tilde{G}_{xx}=0,\\
		(\partial_t^2+|\vec{k}|^2) \tilde{G}_{zy}  +i\dot{\Theta} k_z \tilde{G}_{xy}=0.
	\end{cases}
\end{equation}
We will solve this coupled differential equation system by a perturbative approach.
We expand the formalism up to  second order in $\Theta_0$ in the following way.\\
Consider $G_{xy}$ and $G_{yy}$, whose equations  are coupled between themselves only.
To  order zero we get
\begin{equation}
	\begin{split}
		G_{yy}(\vec{k},t,t') &= H(t-t') \; \times \frac{1}{|\vec{k}|}\, \sin\left[|\vec{k}| (t-t')\right] \label{greensol},\\
		G_{xy}(\vec{k},t,t')&=0,
	\end{split}
\end{equation}
according to the initial conditions, where the function $H(t-t')$ is the usual Heaviside function.
It is then easy to observe that the first-order correction $G^{(1)}_{yy}$ is zero because it satisfies the same equation as $G^{(0)}_{xy}$ with the same boundary conditions. This is not true for $G^{(1)}_{xy}$ because $G^{(0)}_{yy}$ is not zero:
\begin{equation}
	(\partial_t^2+|\vec{k}|^2) G^{(1)}_{xy}+i\dot{\Theta}   k_z G^{(0)}_{yy}=0.
\end{equation}
The solution of this equation has the form of a convolution integral,
\begin{equation}
	\Theta_0 G^{(1)}_{xy}  = i k_z \int_{-\infty}^{\infty} d\omega \;   G_{yy} (t,\omega)  \mathcal{F}(\dot{\Theta}   G^{(0)}_{yy})(\omega),
\end{equation}
where $ \mathcal{F}(\dot{\Theta}   G^{(0)}_{yy})(\omega)$ is the FT of $\dot{\Theta}   G^{(0)}_{yy}$ in $\omega$. It now becomes necessary to know the specific time dependence of the  axion field in order to evaluate the Fourier Transform of the product of the time derivative of the axion field and the Green's function,
\begin{equation}
	\mathcal{F}(\dot{\Theta}   G^{(0)}_{yy})(\omega)= \Theta_0 \omega_a \int dt' e^{i \omega t'}\cos{(\omega_a t')}  G^{(0)}_{yy}(t')=\frac{\Theta_0 \omega_a}{2} \left[\frac{1}{(\omega-\omega_a)^2-|\vec{k}|^2}+\frac{1}{(\omega+\omega_a)^2-|\vec{k}|^2} \right].
\end{equation}
Analogously, we obtain:
\begin{equation}
	\tilde{G}^{(1)}_{xy}  = i |\vec{k}|\frac{\Theta_0 \omega_a}{2}  \frac{1}{\omega^2-|\vec{k}|^2}\left[\frac{1}{(\omega-\omega_a)^2-|\vec{k}|^2}+\frac{1}{(\omega+\omega_a)^2-|\vec{k}|^2} \right],
\end{equation}
	\begin{equation}
	\tilde{G}^{(2)}_{yy}  = \tilde{G}^{(2)}_{xx}=-|\vec{k}|^2 \frac{\Theta_0^2 \omega_a^2}{4}  \frac{1}{\omega^2-|\vec{k}|^2}\left[\frac{1}{(\omega-\omega_a)^2-|\vec{k}|^2}+\frac{1}{(\omega+\omega_a)^2-|\vec{k}|^2} \right]^2.
	\end{equation}
This is a result compatible with what  could have been expected from the HF approximation discussed above.
It is now understandable that according to the results above, a useful ansatz for the  expansion  is
\begin{equation}
	\tilde{G}_{jk}(\vec{k}, \omega) \sim \tilde{G}^{(0)}_{jk}(\vec{k}, \omega)+\tilde{G}^{(1)}_{jk}(\vec{k}, \omega)+\tilde{G}^{(2)}_{jk}(\vec{k}, \omega),
\end{equation}
where
\begin{subequations}
	\begin{align}
		\tilde{G}^{(0)}_{jk}(\vec{k}, \omega)&=\frac{1}{(\omega^2-k^2)} ,\\
		\tilde{G}^{(1)}_{jk}(\vec{k}, \omega)&=\frac{i}{(\omega^2-k^2)^2}\left[\frac{1}{(\omega-\omega_a)^2-|\vec{k}|^2}+\frac{1}{(\omega+\omega_a)^2-|\vec{k}|^2} \right] \epsilon_{jkl}\beta_l,\\
		\tilde{G}^{(2)}_{jk}(\vec{k}, \omega)&=\frac{\beta^2}{(\omega^2-k^2)}\left[\frac{1}{(\omega-\omega_a)^2-|\vec{k}|^2}+\frac{1}{(\omega+\omega_a)^2-|\vec{k}|^2} \right]\left[ \delta_{jk}+\frac{\epsilon_{jkl}\beta_l}{\omega^2-k^2} \right]+\\+&\frac{1}{(\omega^2-k^2)} \left[\frac{1}{(\omega-\omega_a)^2-|\vec{k}|^2}+\frac{1}{(\omega+\omega_a)^2-|\vec{k}|^2} \right]^2\Theta_0^2 \omega_a^2 k_j k_k.
	\end{align}
\end{subequations}
It is worth mentioning that the results for axion oscillating background are compatible with models of axion echo, treated recently in Ref.~\cite{PhysRevLett.123.131804}.
The terms needed in that context can be worked out as done before with a constant time derivative.
\begin{figure}[h]
	\label{coherent}
	\begin{center}
		\includegraphics[width=0.7 \textwidth, keepaspectratio=true]{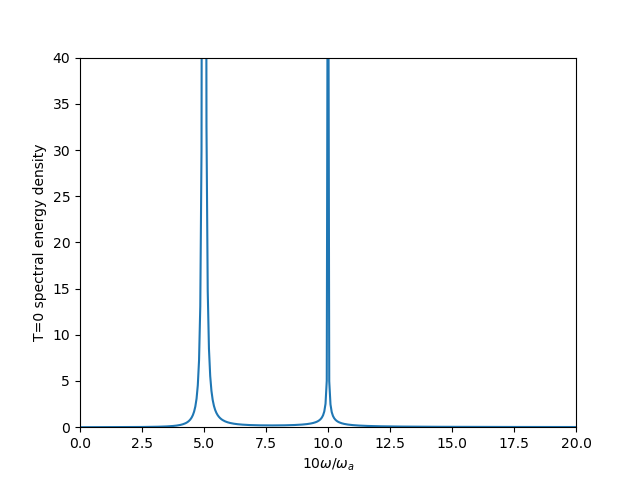}
	\end{center}
	\caption{Graph of the function f in Eq.~(\ref{eqf}) with $\Theta_0=1$.
		 Notice that for $\omega >\omega_a$ we have the expected behaviour from Eq.~(\ref{dothecon}), whose infrared cut-off is the peak at $\omega=\omega_a$, associated with the limiting case of decay of one axion to only one detectable photon. Furthermore, we have a peak at $\omega=\frac{\omega_a}{2}$, a result compatible with the axion echo phenomenon \cite{PhysRevLett.123.131804}, so the physical process behind is the decay of an axion of energy $\omega_a$ to two photons with energy $\frac{\omega_a}{2}$.}
\end{figure}

It is also possible in this case to evaluate the second-order contribution to energy density:
\begin{equation}
\rho^{(2)}(\omega)\; d \omega=d \omega \; \omega^2 \left[ \frac{1}{2} f(\omega,\Theta_0, \omega_a)+ \frac{1}{e^{\omega/T}-1}  f(\omega,\Theta_0, \omega_a)  \right],
\end{equation}
where
\begin{equation}\label{eqf}
f(\omega,\Theta_0, \omega_a)= \frac{ \Theta_0^2 \omega_a^2 \omega (8 \omega^4+\omega_a^4)}{2(\omega+\omega_a)|\omega-\omega_a| (\omega_a^2-4\omega^2)^2}.
\end{equation}
A numerical plot of the function $f$ is displayed in Figure 1: from it one could notice how the validity of an expression of Casimir force between two conducting plates analogous to the case of time increasing axion field is true for $L^{-1} \gg \omega_a$, as mentioned before. For bigger L one appreciates the deviations from that case and consider that the behaviour of the modification to energy density going as $1/\omega$ needs to be corrected with $1/|\omega-\omega_a|$ and has a lower cut-off at $\omega=\omega_a$. Furthermore, one needs to take care of the additional  contribution of "axion echo" \cite{PhysRevLett.123.131804} at $\omega=\frac{\omega_a}{2}$.

This last contribution has a straightforward physical interpretation    :it is associated to the production of virtual photons with frequency $\omega=\frac{\omega_a}{2}$ from the decay of an axion, while the peak at $\omega=\omega_a$ is associated to a decay of one axion to only one detectable photon.
\subsection{A possible way to boost axionic Casimir effect in a a current Universe experimental set-up}
In this subsection we will consider a dielectric system containing two interfaces separating media of refractive indices $n_1$ and $n_2$ inside an  uniform magnetic field $B_e$ and we assume that the media are elastic so that medium 2 can be "turned back" and glued to the left side of medium 1. This is a ring-formed system that we have been treated in Ref.~\cite{PhysRevD.107.043522}, but we extend the theory treated there.
\begin{figure}[h]\label{fig:1}
	\includegraphics[scale=0.3]{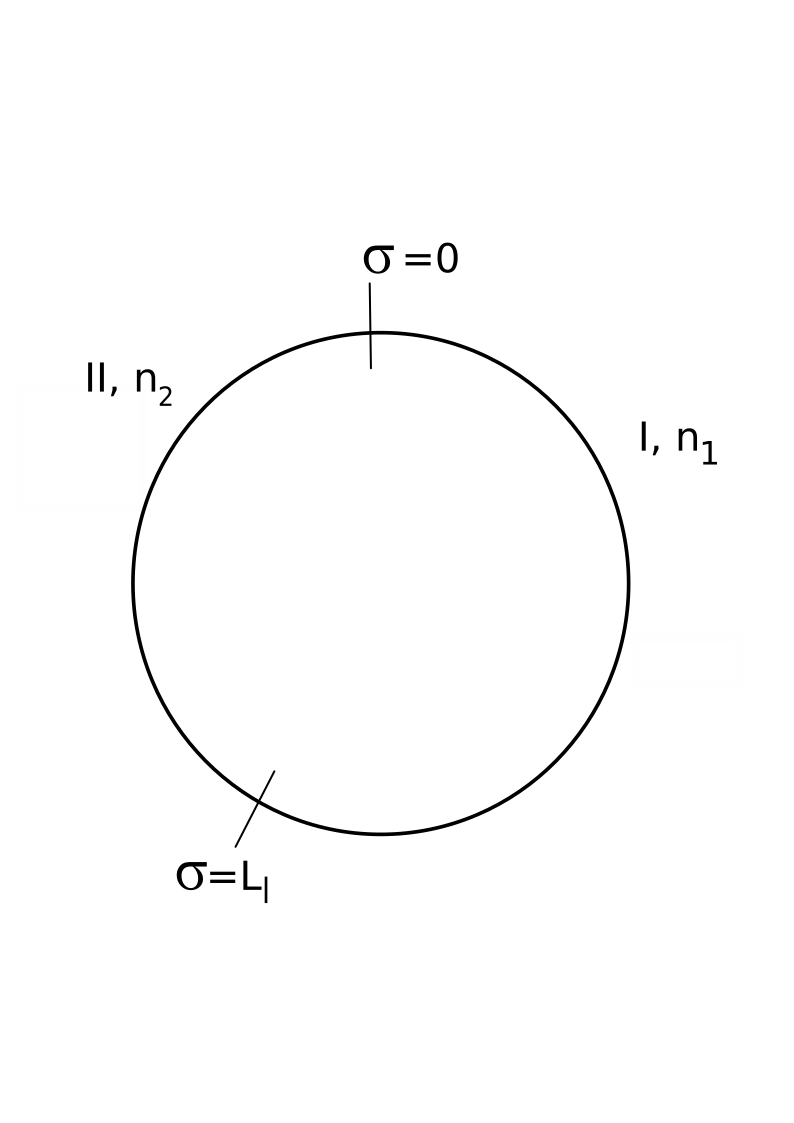}
	\caption{Geometry and notation of the closed string of interest. Figure from Ref.~\cite{PhysRevD.107.043522}}.
\end{figure}
\\Figure 2 shows the configuration and here we treat the most fundamental results from \cite{PhysRevD.107.043522} in order to develop here our new results relative to axion-induced radiation pressure.\\ $\sigma$ is the length coordinate along the string, such that the two junctions are at $\sigma=0$ and $\sigma = L_{I}$. The total length of the string is $L=L_I+L_{II}$ and the junctions $\sigma=0$ and $\sigma=L$ are overlapping.  The string is lying in the $xy$ plane, and a strong uniform magnetic field ${\bf B}_e$ is applied in the $z$ direction.
In Ref.~\cite{PhysRevD.107.043522} we evaluated the stationary oscillations of the electromagnetic oscillations in the string. If $E_I(\sigma,t)$ and $E_{II}(\sigma,t)$ are the electric fields in the two regions, we have in complex representation
\begin{subequations}
	\begin{align}
		E_I(\sigma,t)=\xi_Ie^{in_1\omega \sigma-i\omega t}+\eta_Ie^{-in_1\omega \sigma -i\omega t},\\
		E_{II}(\sigma,t)=\xi_{II}e^{in_2\omega(\sigma-L_I)-i\omega t}+ \eta_{II}e^{ -in_2\omega (\sigma-L_I)-i\omega t},
	\end{align}
\end{subequations}
where $\xi_I, \eta_I,\xi_{II},\eta_{II}$ are constants. Analogously, using the relationship $H= nE$, one gets the magnetic fields.
One could treat such a system by considering a generic $\omega$, however, as we discussed in Ref.~\cite{PhysRevD.107.043522}, the axion-generated oscillations would be suppressed by a factor $\delta(\omega-\omega_a)$, so the only relevant frequency is $\omega_a$ for our purposes. Then, we take $\omega=\omega_a$ in the following.
By imposing the boundary conditions at the junctions to be closed,  we found it for particular cases and here we deal with a particular case of interest,  namely when $n_2$ becomes large in comparison to $n_1$ so that the ratio $x\rightarrow 0$ and we consider the case where the length $L_I\rightarrow 0$, corresponding to a kind of point defect sitting on an otherwise uniform string. As $\delta_1\rightarrow 0$ in this case, we see  that
\begin{equation}
	\xi_I= \frac{E_0}{2\varepsilon_1}, \quad x\rightarrow 0, L_I \rightarrow 0,
\end{equation}
which is a real quantity.
We also have for this last case:
\begin{align}
	\eta_I=\xi_I,\\
	\eta_{II}=\xi_{II}=0.
\end{align}
We have then an electromagnetic energy density $u_{1def}$ inside the point defect:
\begin{equation}
	u_{1def}=\frac{E_0^2}{4 \epsilon_1}.
\end{equation}
It is then worth mentioning the similarities and differences with the system of a single dielectric plate treated by Millar et al. \cite{Millar_2017}.
Indeed our system is like the one in \cite{Millar_2017}, but when the two outward dielectric media are closed to form a ring.
That suggestes the possibility that the case of a closed ring with $N$ multiple dielectric plates could be of interest, since we could expect to have an enhancement of $ \sim N^2$ for $u_{1def}$ as in the setup of \cite{Millar_2017}.

The treatment of a ring with multiple regions (e.g. multiple point defects) is surely very complicated analitically and would need a numerical treatment. This can be subject of future works, along with other possible geometries, such a cylindrical one with multiple conducting or dielectric plates, as mentioned in \cite{PhysRevD.107.043522}.
Here we just give an order of magnitude of the expected radiation pressure $p$. If we consider the behaviour of $p$ going as $\sim N^2$, we have:
\begin{equation}
	p \sim N^2\frac{E_0^2}{4 \epsilon_1}.
\end{equation}
If we take from the system in Ref.~\cite{Millar_2017} with the following orientative values  $B_e \sim \SI{10}{T}$, $\Theta_0 \sim 10^{-21}$, $\epsilon_1 \sim 1$, $N \sim 80$, one gets:
\begin{equation}
	p \sim  3.2 \times 10^{-13} \, \frac{1}{\epsilon_1} \, \left(\frac{\Theta_0}{10^{-21}}\right)^2 \, \left(\frac{N}{80}\right)^2\left(\frac{B_e}{\SI{10}{T}}\right)^2 \, \SI{}{Pa} \, .
\end{equation}
It is worth noticing that for the orientative values we get a pressure that is just one order of magnitude lower than the lowest pressures gotten in laboratory (see e.g. Refs.~\cite{thompson1977characteristics,redhead1999extreme}), but it would be possible to get or overcome those values of pressure with a bigger number $N$.
\section{Optical properties of a toy model and axion domain walls}\label{wavedomain}
The purpose of this subsection is to study the propagation of an electromagnetic wave in the presence of a toy model for the axion domain wall. The toy wall is taken to be a planar sheet in which the axion field increases linearly with the longitudinal coordinate $z$ between $z=0$ and $z=L$ and it is constant and uniform outside. We assume a static wall, so $a=a(z)$ only, thus no time dependence.
As mentioned above, we first consider a space-dependent axion field with planar symmetry:
\begin{equation}\label{toy}
\Theta(z)=\begin{cases}
		0  \; \; \; \textit{if} \; \; \; z < 0, \\
		\frac{\Theta_0}{L}z \; \; \; \textit{if} \; \; \; 0< z < L,\\
		\Theta_0 \; \; \; \textit{if} \; \; \; z>L.
	\end{cases}
\end{equation}
The model was treated earlier in Ref.~\cite{universe7050133}: we will here make some corrections.
It serves as a toy model for  'localized' axion configurations since  space-dependent axion configurations have typical lenghts $L \sim m_a^{-1}$.
\begin{figure}[h] \label{axion-domain-wall}
	\begin{center}
		\includegraphics[width=0.9 \textwidth, keepaspectratio=true]{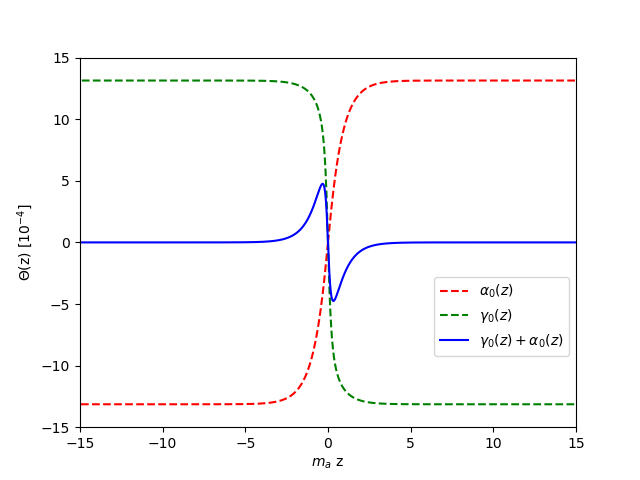}
	\end{center}
	\caption{Explicit graph of the static axion domain wall configuration $\Theta(z)=g_{a \gamma \gamma} a(z)$ where $a(z)=\gamma_0(z)+\alpha_0(z)$ is reported in solid line in blue, evaluated as the ones in Ref.~\cite{PhysRevD.32.1560}.We reported also the correspondent values for $\gamma_0(z)$ and $\alpha_0(z)$.}
\end{figure}
We will now investigate the axion modifications of the electromagnetic wave in the system. We mention in passing that possible other approximation could be:
\begin{equation}\label{tanh}
	\Theta(z)=\frac{\Theta_0}{2} \left[\tanh\left(\frac{z}{L} \right)-1 \right].
\end{equation}
As shown in Ref.~\cite{PhysRevD.32.1560} the 'effective field' in Axion-Modified Maxwell equations for an actual cosmological static axion domain wall gets two contributions: one $\alpha_0(z)$ from the phase of Peccei-Quinn field, i.e. the axion field and $\gamma_0(z)$ from the neutral pion field, since the pion has an interaction term to electromagnetic field analogous to Eq.~(\ref{eq0}).
The two contributions are related because of the equilibrium condition for a static wall:
\begin{equation}
	\tan{\gamma_0}=-\xi \tan{\alpha_0},
\end{equation}
where $
\xi=\frac{m_u-m_d}{m_u+m_d}$, where $m_u$ is the mass of the up quark and $m_d$ is the mass of down quark.
In our numerical evaluation here we used the value $\xi=0.3$ as done by Ref.~\cite{PhysRevD.32.1560} and we evaluated $\alpha_0(z)$ by solving numerically the equation:
\begin{equation}\label{eqalpha}
\frac{1}{m_a^2}\frac{d^2 \alpha_0}{dz^2}=\frac{\sin{\alpha_0} \cos{\alpha_0}}{\sqrt{\cos^2{\alpha_0}+ \xi^2 \sin^2{\alpha_0} }},
\end{equation}
obtained by the same Ref.~\cite{PhysRevD.32.1560}.
We plot one of the possible solutions of Eq.~(\ref{eqalpha}) in Figure 3. It is worth noticing that it is a 'localized' axion configuration whose typical lenght is of the order of $m_a^{-1}$.
\subsection{Exact calculations for toy model}\label{exactcal-toy}
As above, we consider  an axion configuration that is   dependent on $z$ only.  At first, we assume $\beta(z) = \partial_z \Theta(z)$   to be an arbitrary function of $z$.
The incident electromagnetic wave is propagating along the $z$-axis and is transversely polarized  (the same configuration was assumed in  Ref.~\cite{PhysRevD.32.1560}).
This configuration is useful because a domain wall  is invariant under Lorentz boosts parallel to the wall surface, so it is always possible to find a reference frame where the wave is incident normally, as also noticed in the same \cite{PhysRevD.32.1560}.\\
This simplifies the calculations to the first order  in the 4-potential because the effective axion charge density is zero.
We then consider the following equation:
\begin{equation}\label{eqHeav}
	\Box \vec{A}=-\nabla \Theta \wedge \frac{\partial \vec{A}}{\partial t},
\end{equation}
and  search for an exact solution.
We calculate the FT of this equation in time $t$ and in the $x$  and $y$ coordinates,  and  get the following expressions for the FT components of the vector potential:
\begin{equation}\label{fracco}
	\begin{split}
		(-\partial_z^2-\kappa^2)\mathcal{A}_x&=i \omega \beta(z) \mathcal{A}_y, \\
		(-\partial_z^2-\kappa^2)\mathcal{A}_y&=-i \omega \beta(z) \mathcal{A}_x, \\
		(-\partial_z^2-\kappa^2)\mathcal{A}_z&=0.
	\end{split}
\end{equation}
When $k_x=k_y=0$, we have $\kappa=\omega$.
Equations (\ref{fracco}) imply that the gauge fields defined as $\mathcal{A}_{\pm}=\mathcal{A}_x \pm i \mathcal{A}_y$  satisfy:
\begin{equation}\label{fracco1}
	\begin{split}
		(-\partial_z^2-\kappa^2)\mathcal{A}_{\pm}&=\pm  \omega \beta(z) \mathcal{A}_{\pm}.
	\end{split}
\end{equation}
From now on we assume to deal with the toy model configuration (\ref{toy}), so $\beta(z)=0$ outside the wall, while $\beta(z)=\frac{\Theta_0}{L}=\beta_z$ inside the wall.
 We notice that the axion configuration (\ref{toy}) has not well-defined values of the z derivative at $z=0,L$ since the derivative has a Heaviside behaviour in such points. However, this is not problematic for treating Eq.~(\ref{eqHeav}) since it just tells us that the second z-derivative of vector potential has a Heaviside behaviour, but vector potential and its first z-derivative are continuous.

Indeed, in the following calculations we basically solve Eq.~(\ref{eqHeav}) for the regions $z<0$, $z>L$ and $0<z<L$ (where $\beta(z)$ is well-defined) and then impose continuity of vector potential and its first z-derivative.
This last procedure would be also good in the case one considers axion configurations like (\ref{tanh}) where there would not be Heaviside-like behaviours.

\begin{figure}\centering
	\includegraphics[scale=0.5]{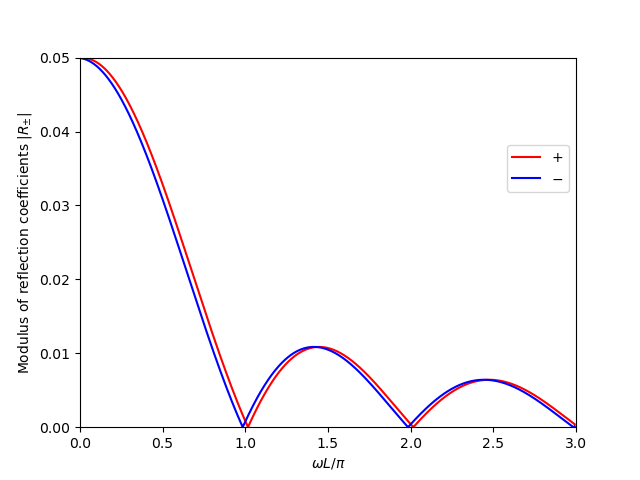}
	\caption{Plot of reflection coefficient modulus for + (red) and - (blue)polarizations.}
	\label{Plotto}
\end{figure}

\begin{figure}\centering
	\includegraphics[scale=0.4]{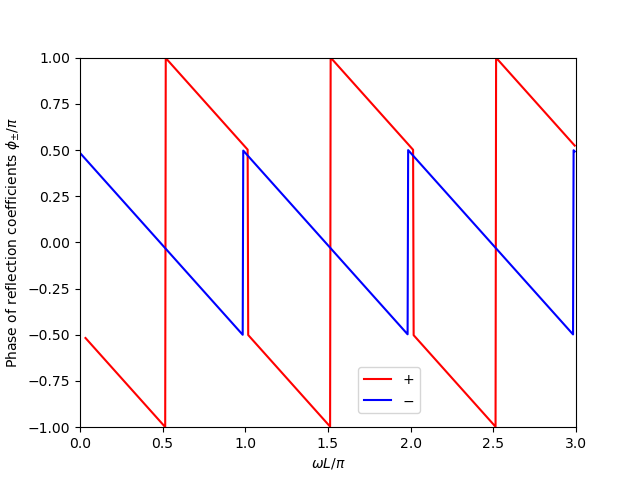}
	\caption{Plot of reflection coefficient phase for + (orange) and - (blue) polarizations.}
	\label{Plotto1}
\end{figure}

Moreover, we consider a plane wave propagating along z.
Then, the equations (\ref{fracco1}) can be interpreted as follows.
For the field $\mathcal{A}_{-}$ we have a usual harmonic oscillator with frequency $\alpha_{-}=\sqrt{\omega^2+ \omega \beta(z)}$ inside (and outside) the   domain wall, while for the field $\mathcal{A}_{+}$ we have a damped wave inside the   domain wall if $\omega> \beta_z$.
The gauge fields $\mathcal{A}_{\pm}$ correspond to left/right circular polarizations, so the basic properties are consequently different for a left circular polarized wave and a right circular polarized one.
We now evaluate the reflected component of the incident orthogonal wave. Recall the geometry: the leftmost region is $z<0$, the slab region is $0<z<L$, and the rightmost region is $z>L$. The continuity conditions at the walls are that the  $A_{\pm}$ fields as well as their $z$ derivatives are continuous. Thus in the leftmost region,
\begin{equation}
	\mathcal{A}_{\pm}(z)_{\text{left}}=A_{\pm}e^{-i \kappa z}+B_{\pm}e^{i \kappa z},
\end{equation}
where the $B$ term is the reflected wave. The wave transmitted outside the slab is
\begin{equation}
	\mathcal{A}_{\pm}(z)_{\text{right}}=C_{\pm}e^{-i \kappa z}.
\end{equation}
To simplify the notation, we write $\alpha$ instead of $\alpha_{\pm}$, and we consider $\alpha$ to be the frequency inside the domain wall:
\begin{equation}
	A_{\pm}(z)_{\text{in}}=A'_{\pm}e^{-i \alpha z}+B'_{\pm}e^{i \alpha z}.
\end{equation}
where $A'_{\pm}$ is the wave transmitted inside the wall.
In order to evaluate the reflection coefficient (defined as $R_{\pm}=\frac{B_{\pm}}{A_{\pm}}) $ and the internal transmission coefficient (defined as $T_{\pm}=\frac{A'_{\pm}}{A_{\pm}}$), we need to impose the continuity conditions on the surfaces $z=0$ and $z=L$:
\begin{equation}
	\begin{cases}
		A+B=A'+B' & \text{Continuity of $\mathcal{A}_{\pm} $ at z=0},\\
		\omega A-\omega B=\alpha A'-\alpha B' & \text{Continuity of first derivatives of $\mathcal{A}_{\pm} $ at z=0},\\
		e^{-i \alpha L} A'+e^{i \alpha L} B'=C e^{-i \omega L} & \text{Continuity of $\mathcal{A}_{\pm} $ at z=L}, \\
		-i \alpha e^{-i \alpha L} A'+i \alpha e^{i \alpha L} B'=-i \omega C e^{-i \omega L} & \text{Continuity of first derivatives of $\mathcal{A}_{\pm} $ at z=L}.
	\end{cases}
\end{equation}
From it we get
\begin{subequations}\label{reflecto}
	\begin{align}
		T &= \frac{2 e^{2 i\alpha L} \omega (\alpha + \omega)}{(-1 + e^{2 i \alpha L}) \alpha^2 + (-1 + e^{2 i \alpha L}) \omega^2 + 2 (1 +e^{2 i \alpha L}) \alpha \omega}, \\
		R &= -\frac{(-1 + e^{2 i \alpha L}) (\alpha^2 - \omega^2)}{(-1 + e^{2 i \alpha L}) \alpha^2
			+ (-1 + e^{2 i \alpha L}) \omega^2 + 2 (1 + e^{2 i \alpha L}) \alpha \omega}.
	\end{align}
\end{subequations}
Some fundamental properties of the reflection coefficient are shown in Figs. \ref{Plotto} and \ref{Plotto1}. Modules and phases are given, where we have taken the unit frequency to be $1/L$ and  $\Theta_0=10^{-3}$.
We notice that there is most reflection, for both modes,  when $\omega=0$ and $\omega=\frac{\Theta_0}{L}$ but there is still reflection at higher frequencies at intervals of $\omega=\frac{\Theta_0}{L}$.
\subsection{Dispersion relation}\label{Casimirdisperdo}
In the previous section we obtained the following dispersion relation for the toy model (\ref{toy}):
\begin{equation}
	\omega^2-k_{z}^2=\mp \omega \beta(z),
\end{equation}
implying
\begin{equation}
	\omega_{\pm}=\sqrt{ \left(k_z^2+ \frac{1}{4}\beta^2(z) \right)} \pm \frac{1}{2}\beta(z).
\end{equation}
This is the same dispersion relation as obtained in  \cite{universe7050133}. However, this is not the general dispersion relation because, as  discussed in the previous section, it applies to a single linear polarization only in its reference frame where the wave vector lies along z direction. Applying an inverse Lorentz boost (so in x and y directions) we can generalize the relation to take the following form, similar to the one obtained in Ref.~\cite{PhysRevD.100.045013}:
\begin{equation}
	\omega=\sqrt{k_x^2+k_y^2+\left(\sqrt{ \left(k_z^2+ \frac{1}{4}\beta^2_z \right)} \pm \frac{1}{2}\beta_z \right)^2   }.
\end{equation}
\section{Space-dependent axion field. Casimir aspects}\label{spazio}
We now consider the simplest case of a planar symmetry  in which the axion field is constant in time but depends on the longitudinal coordinate z only. The gradient of the
field is thus directed along the $z$-axis; we will denote it $\nabla_z \Theta=\beta(z)$.
We get then the following equations for the reduced Green's function in component form:
\begin{subequations}\label{pippogxxl}
	\begin{align}
		-(\partial_z^2+\omega^2-k_y^2) g_{xx}+ik_x(\partial_z g_{zx}+ik_y g_{yx})-i \omega \beta(z) g_{yx} =-\delta(z-z'), \\
		-(\partial_z^2+\omega^2-k_x^2) g_{yy}+ik_y(\partial_z g_{zy}+ik_x g_{xy})+i \omega \beta(z) g_{xy} =-\delta(z-z'), \\
		-\kappa^2 g_{zz}+\partial_z(ik_x g_{xz}+ik_y g_{yz}) =-\delta(z-z'),
	\end{align}
\end{subequations}
and
\begin{subequations}\label{pippogzyl}
	\begin{align}
		-(\partial_z^2+\omega^2-k_x^2) g_{yx}+ik_y(ik_x g_{xx}+\partial_z g_{zx})+i \omega \beta(z) g_{xx} =0, \\
		-(\partial_z^2+\omega^2-k_y^2) g_{xy}+ik_x(ik_y g_{yy}+\partial_z g_{zy})- i \omega \beta(z) g_{yy}  =0 \\
		-(\partial_z^2+\omega^2-k_y^2) g_{xz}++ik_x(ik_y g_{yz}+\partial_z g_{zz})=0,\\
		-\kappa^2 g_{z i}+\partial_z(ik_y g_{yi}+ik_x g_{xi} )=0 \text{ if $i \neq z$},
	\end{align}
\end{subequations}
as well as  other analogous expressions.
We here develop briefly the case of the domain wall.

\subsection{Casimir force on domain wall}\label{Casimir-dm}
We can manipulate the equations (\ref{pippogxxl}) and (\ref{pippogzyl}) and the other equations by some preliminar manipulations.
Combining (\ref{pippogxxl} a),(\ref{pippogzyl} a) and (\ref{pippogzyl}) we can get the two following  coupled equations for $g_{xx} $ and $g_{yy}$:
\begin{equation}
	\begin{split}
		\omega^2 \left(\partial_z^2+\kappa^2+i \beta(z) \frac{k_x k_y}{\omega} \right)g_{xx}+ i \omega \beta(z)  (\omega^2-k_x^2) g_{yx}&= (\omega^2-k_x^2) \delta(z-z'),\\
		\omega^2 \left(\partial_z^2+\kappa^2-i \beta(z) \frac{k_x k_y}{\omega} \right)g_{yx}-i \omega \beta(z)  (\omega^2-k_y^2) g_{xx}&= -k_x k_y \delta(z-z').
	\end{split}
\end{equation}
It is convenient to define $ g^e_{ij}=-\omega^2 g_{ij} $  and we notice it is also convenient to define the functions
\begin{equation}
	g_{\pm}(z,z')=g^e_{xx}(z,z') + \gamma_{\pm}(\tilde{k},\omega) g^{e}_{yx}(z,z'),
\end{equation}
where
\begin{equation}
	\gamma_{\pm}(\tilde{k},\omega)=\frac{k_x k_y \pm i \omega \kappa}{\omega^2-k_y^2}.
\end{equation}
We then get the following simpler equations:
\begin{equation}\label{equpm}
	\begin{split}
		\left[ \partial_z^2+\kappa^2 \pm  \beta(z) \kappa \right]g_{\pm}= \left[\frac{\omega^2 \kappa^2}{\omega^2-k_y^2} \mp \frac{i \omega \kappa k_x k_y}{\omega^2-k_y^2} \right] \delta(z-z').
	\end{split}
\end{equation}
From solving Eq.~(\ref{equpm}) it is then straightforward to find $g_{xx}$ and $g_{yx}$ and from these two can find $g_{zx}$.\\
It is also easy to find an analogous\footnote{It is trivial that it is analogous because the system is invariant under rotation around $z$-axis} result with $g_{yy}$, $g_{xy}$ and $g_{zy}$:
\begin{equation}
	\begin{split}
		\left[ \partial_z^2+\kappa^2 \pm  \beta(z) \kappa \right]\tilde{g}_{\pm}= \left[\frac{\omega^2 \kappa^2}{\omega^2-k_x^2} \mp \frac{i \omega \kappa k_x k_y}{\omega^2-k_x^2} \right] \delta(z-z'),
	\end{split}
\end{equation}
where
\begin{equation}
	\tilde{g}_{\pm}(z,z')=g^e_{yy}(z,z') + \tilde{\gamma}_{\pm}(\tilde{k},\omega) g^{e}_{xy}(z,z').
\end{equation}
We notice then that for expressions such as $\tilde{g}_{\pm}(z,z')$ and 	$g_{\pm}(z,z')$  we need from the continuity of Green's function and their equation to have for $z,z'<0$:
\begin{equation}
g_{\pm}(z,z') \propto \frac{1}{2 \kappa} (e^{-i \kappa |z-z'|}+ R_{\pm}(\kappa)e^{i \kappa (z+z')}),
\end{equation}
where $R_{\pm}(\kappa)$ are exactly the reflection coefficients (\ref{reflecto}).
Then, if we calculate the zz component of Energy-momentum tensor by expression (\ref{Tzz}), we can get after a lot of calculations the temperature-dependent Casimir force per unit area:
\begin{equation}
	f(T,L)=-T \sum_{\pm} \sum_m^{+ \infty \, '} \int_{\zeta_m}^{+ \infty} \kappa^2 d\kappa \, \frac{|R_{\pm}(\kappa)|^2 e^{-2L\sqrt{q^2 \pm q \beta_z}}}{1-|R_{\pm}(\kappa)|^2 e^{-2 L \sqrt{q^2 \pm q \beta_z}}}.
\end{equation}
In the following Subsection, we discuss its high temperature limit ($L^{-1} \ll T$), that can be obtained by considering the first term $m=0$ in the sum and it is equal to:
\begin{equation}\label{fo}
	\begin{split}
		f(T,L) = 2 \int \frac{d^3k}{(2 \pi)^3} \left[R^2_{+}(k_z)+R^2_{-}(k_z)  \right] \frac{k^2_z}{\omega} \frac{1}{e^{\beta \omega}-1} H(k_z)=\\=2 \int \frac{d^2 k}{(2 \pi)^3} \int_{0}^{+\infty} dk_z \left[R^2_{+}(k_z)+R^2_{-}(k_z)  \right] \frac{k^2_z}{\omega} \frac{1}{e^{\beta \omega}-1} ,
	\end{split}
\end{equation}
It can be demonstrated by calculating the last integral in the variables $k_x$ and $k_y$.
However, it is very hard to develop the calculations with the exact explicit expressions (\ref{reflecto}) of  $R^2_{+}(\kappa) $.
In the following, we approximate the reflection coefficients $R_{\pm}(\kappa)$ in a similar way as in Ref.~\cite{PhysRevD.32.1560} and also based on our results for the toy model.
For simplicity we define the useful quantity $m_L=\frac{\Theta_0}{L}$.

The reflection coefficients can be approximated to be $R^2_{+}(k_z) \sim \zeta m_L \, \delta(k_z)$ and $R^2_{-}(k_z) \sim \zeta \, m_L \, \left[\delta(k_z+m_L)+\delta(k_z-m_L)\right]$ where $\zeta$ is a numerical factor such that fits better the behaviour of the reflection coefficients.
This can be justified by noticing that, although the exact behaviours of the square modula of $R_{\pm}(k_z) $  are complicated, it is clear that a crucial frequency for our system is the same $m_L$, and  by the graphs in Figures~\ref{Plotto} and \ref{Plotto1} and a qualitative description of them, they can be approximated to be $R^2_{+}(k_z)\sim  \frac{\zeta}{\pi} \Theta_0^2  m_L^2 \frac{\sin^2{(k_z/m_L)}}{k^2_z}$ and  $R^2_{-}(k_z) \sim \frac{\zeta}{\pi} \Theta_0^2  \frac{\sin^2{(k_z/m_L-1)}}{(k_z/m_L-1)^2}$.
Then one exploits the distributional relation:
\begin{equation}
	\frac{\sin^2{(\epsilon t)}}{t^2} \sim \pi \epsilon \delta(t)  \qquad \text{for} \qquad  \epsilon \rightarrow 0,
\end{equation}
and we can get the expressions if $\epsilon=( m_L)^{-1}$.

The use of the Dirac delta limit is only useful for simplifying the calculations, however, an exact calculation would correspond to taking care of the secondary peaks of the graphs in Figure~\ref{Plotto}  along with the detailed structure of them and also of the main peak (the last one is namely $k_z=0$ for $R_{-}(k_z)$ and $k_z=m_L$ for $R_{+}(k_z)$)    just modify the numerical factor $\zeta$.
The Dirac delta approximation then takes care of the main contribution coming from the main peak of Figure~\ref{Plotto}, approximating it as a strict peak.

\subsection{Evaluation of the temperature-dependent electromagnetic pressure on domain wall}\label{dio}
\begin{figure}[h]\label{figuredominio}
	\includegraphics[scale=0.5]{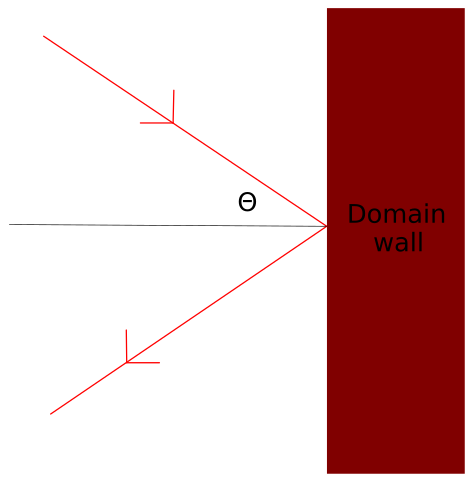}
	\caption{Visual representation useful for the calculation of radiation pressure acting on an axion domain wall}
\end{figure}
As we mentioned previously, the calculation of the Temperature dependent Casimir pressure acting on an axion domain wall is not immediate.
We can anyway get a first physical idea by getting the expression (\ref{fo}), making a calculation similar to the one made by Huang and Sikivie \cite{PhysRevD.32.1560} and to the one made in Ref.~\cite{Blasi_2023}, that also highlights the physical meaning and let us extend the result to non-static walls.\\
Our results can be first applied to QCD axion domain walls, but also to ALP domain walls \cite{Blasi_2023} and overall strong topological insulators \cite{PhysRevD.100.045013, nenno2020axion,yan2021majorana}.
Namely, the last cited application can be the same system treated in Ref.~\cite{PhysRevD.100.045013}, if we consider the toy model configuration for the axion field (\ref{toy}), but without the presence of two conducting plates, so the comparison between our results and the ones in the cited reference is straightforward.

The idea is basically to calculate kinetically the radiation pressure acting on the toy model (\ref{toy}) at temperature $T$, whose results can be also indicative for cosmological walls, as we discussed before.
If we have a circularly polarized electromagnetic wave incident on the axion domain wall this is partly reflected as discussed in Section \ref{wavedomain}.
We refer to Figure~\ref{Plotto1} for the geometry of the system.

If the incident wave (that can be left-circularly or right-circularly polarized) has momentum density that is equal to its energy density $u_{em,\pm}(\omega)$ it is partly reflected back with momentum density that is equal to $R^2_{\pm}(k_z) \, u_{em,\pm}(\omega)$ and partly transmitted with momentum density equal to $T^2_{\pm}(k_z) \, u_{em,\pm}(\omega)$, depending on its polarization.

We preliminarly observe that, as also visible from what gotten for the toy model, the reflection and transmission coefficients can only depend on $k_z$, since the motion parallel to the domain wall cannot affect the dynamics as the domain wall is invariant for parallel boosts.
It then experiences a variation of momentum equal to $ \delta t \, \delta A \, \left[2 R^2_{\pm}(k_z) \cos{\theta}^2 \right] u_{em,\pm}(\omega)$ where $\delta A$ is the differential area and $\delta t$ is the differential time. That expression is consequence of the relation $R^2_{\pm}(k_z)+T^2_{\pm}(k_z)=1$, so $1+R^2_{\pm}(k_z)-T^2_{\pm}(k_z)=2R^2_{\pm}(k_z)$.
If we divide by $\delta A$ and $\delta t$ we get the pressure exerted from that mode with that specific polarization.

Summing on the two polarization and all frequencies, we get the following  pressure $\mathcal{P}_L$ acting on the left of the static wall, similarly to what is obtained in Ref.~\cite{Blasi_2023}:
\begin{equation}\label{pressure}
	\begin{split}
	\mathcal{P}_L = \mathcal{P}= 2 \int \frac{d^3k}{(2 \pi)^3} \left[R^2_{+}(k_z)+R^2_{-}(k_z)  \right] \frac{k^2_z}{\omega} \frac{1}{e^{\beta \omega}-1} H(k_z)=\\=2 \int \frac{d^2 k}{(2 \pi)^3} \int_{0}^{+\infty} dk_z \left[R^2_{+}(k_z)+R^2_{-}(k_z)  \right] \frac{k^2_z}{\omega} \frac{1}{e^{\beta \omega}-1} ,
	\end{split}
\end{equation}
which is equal to the pressure $\mathcal{P}_R$ acting on the right of the wall because a static domain wall is left-right symmetric. $H$ is the Heaviside function, taking care of just the contribution of photons coming from left  and here $\omega=\sqrt{k^2_x+k^2_y+k^2_z}$, so we exploited the relation $k_z= \omega \cos{\theta}$.

The calculation of $\mathcal{P}$ is not trivial because it is dependent on the details of the axion configuration, from which reflection coefficients depend. For the case of the toy model we can evaluate it by using the obtained explicit expressions. 
In this way, the integral (\ref{pressure}) becomes explicitly as:
\begin{equation}
	\mathcal{P}_L=2  m_L^3\int \frac{d^2 k}{(2 \pi)^3} \frac{1}{\sqrt{k^2+m_L^2}} \frac{1}{e^{\beta \sqrt{k^2+m_L^2}}-1},
\end{equation}
since one can use the parity property of the integrand under the change of variable $k_z \rightarrow -k_z$ and the defining property of the Dirac delta $\int_{-\infty}^{\infty} dx f(x) \, \delta(x-x_0)=f(x_0)$.
This integral can be evaluated by noticing that the integrand is dependent on the variable $k=\sqrt{k_x^2+k_y^2}$, on the physical quantities $m_L$ and $\beta$, so we can use polar coordinates and making the substitution $k_i \rightarrow k_i/m_L$ to get an integral just dependent on $\beta m_L$:
\begin{equation}
	\mathcal{P}_L= \frac{m_L^4}{  \pi^2} \int_0^{+\infty} d k' \,  \frac{k'}{\sqrt{k'^2+1}} \frac{1}{e^{\beta m_L \sqrt{k'^2+1}}-1}.
\end{equation}
This integral can be evaluated by noticing that a primitive function (indefinite) integral $F(x)$ of the function
\begin{equation}
	f(x)=\frac{x}{\sqrt{x^2+1}} \frac{1}{e^{\alpha \sqrt{x^2+1}}-1}
\end{equation}
is trivially the function $F(x)=\alpha^{-1} \ln{(e^{\alpha \sqrt{x^2+1}}-1)}-\sqrt{x^2+1}$.
Consequently, it can be easily found that
\begin{equation}\label{solo0}
		\mathcal{P}= \frac{m_L^4 }{  \pi^2}  \left[1-\frac{T}{m_L} \ln{(e^{\beta m_L}-1)}\right].
\end{equation}
We get then expressions agreeing partly with those mentioned in Ref.~\cite{PhysRevD.32.1560}:
\begin{subequations}\label{Pl0}
	\begin{align}
	\mathcal{P}= \frac{m_L^3}{\pi^2} \,  \, T e^{-\frac{m_L}{T}} \, \, \, \, \, \, \, \, \, \,   \text{for $T \ll m_L$},\label{Pl0a}\\
	\mathcal{P}=  \frac{m_L^3  }{\pi^2}\, \, T \, \ln{\left(\frac{T}{m_L}\right)}  \, \, \, \, \, \, \, \, \, \text{for $T \gg m_L$}.\label{Pl0b}
	\end{align}
\end{subequations}
The first limiting case (\ref{Pl0a})  can be found by adopting the asymptotic expression $\ln{(x-1)} \sim \ln{x}-\frac{1}{x}$ for $x \rightarrow +\infty$, where $x$ is in our case $e^{\beta m_L}$.
The second case (\ref{Pl0b}) can be obtained by using  $\ln{(e^x-1)} =\ln{x}+\frac{x}{2}+ \mathcal{O}(x^4)$ for $x \rightarrow 0$ and leaving the more relevant term for $T \gg m_L$, i.e. the logarithm $\ln{x}$.\\
We notice how our result (\ref{Pl0b}) is different from that claimed in Ref.~\cite{PhysRevD.32.1560}, since here we get a behaviour going like $ \sim m_L^3 T \ln(T/m_L)$, that is proportional to $\Theta_0^3$, which is different from a behaviour $\sim \Theta_0^2 m_a^2 T^2$. This is a relevant difference and it is a consequence of the different dependence on temperature $T$ and the observation of the relevance of the threeshold frequency $m_L$, instead of $L^{-1}$, where the reflection coefficients are significant.

It is also of interest to treat the case of a domain wall moving at constant velocity $v$, namely along the direction of the $z$-axis, relatively to the reference frame where the electromagnetic radiation emission is an isotropic blackbody one at temperature $T$.
In such a case, expressions for $\mathcal{P}_L$ and $\mathcal{P}_R$ are similar to the ones in the static case $v=0$, but there are two main differences. The first one is that we do not expect $\mathcal{P}_L$ and $\mathcal{P}_R$ to be equal since the left-right symmetry is broken in such a case, while the second one relies on the need of taking care of the Doppler effect.
Consequently we get, similarly to Ref.~\cite{Blasi_2023},
\begin{equation}
	\mathcal{P}_{L,R}= \frac{m_L^4}{  \pi^2} \int_0^{+\infty} d k' \,  \frac{k'}{\sqrt{k'^2+1}} \frac{1}{e^{ \gamma(v)\beta m_L (\sqrt{k'^2+1}\pm v)}-1}.
\end{equation}
where the $+$ is valid for $\mathcal{P}_{L}$ while $-$ for $\mathcal{P}_{R}$.
This integral can be calculated by noticing that the integrand has a primitive function (indefinite integral) that is analogous to the one of static case. One then gets the exact solution:
\begin{equation}
	\mathcal{P}_{L,R}= \frac{1}{  \pi^2} m_L^4 e^{ \mp \gamma(v) \beta m_L v} \left[1-\frac{T}{m_L} \ln{(e^{\beta m_L(1 \pm v)}-1)}\right].
\end{equation}
We highlight, as mentioned before, that the pressures in the limit $\beta m_a \gg 1$ can be evaluated by just substituting the Bose-Einstein distribution with the limiting Boltzmann factor $e^{ -\gamma(v)\beta m_a (\sqrt{k'^2+1} \pm v)}$, which is indeed a good approximation for $\beta m_a \gg 1$,  since it is surely much bigger than 1. This substitution is also true in the ultrarelativistic limit $v \simeq 1$ for the same reason.
We have for such limiting cases ($\beta m_a \gg 1$ and/or $v \simeq 1$):
\begin{equation}
	\mathcal{P}_{L,R} \propto m_L^3 T \sqrt{1-v^2} e^{- \gamma(v)\beta m_L(1 \pm v)}
\end{equation}
The non-relativistic regime $v \ll 1$ can be well approximated by the solution (\ref{solo0})  with $v=0$.
\subsubsection{Application to Axion Cosmology}
In the case of interest for Axion cosmology $T \gg m_L$, if we consider a static wall, we adopt the expression (\ref{Pl0b}) for such a case.
This result gives us a pressure:
\begin{equation}\label{Pem0}
	P_{em}^0 \sim 10^{-27}\left(\frac{\Theta_0}{10^{-3}}\right)^3   \left(\frac{m_a}{10^{-2} \, \SI{}{eV}}\right)^3  \left(\frac{T}{ \SI{1}{GeV}}\right) \left[1+\ln{\left(10^{-14}\frac{T}{m_a} \right)}   \right]\, \SI{}{MeV}^4.
\end{equation}
This very low pressure can be compared for example to gravitational pressure.
Indeed, according to Ref.~\cite{PhysRevD.32.1560} in the $N_{DW}=1$ case,  a closed axion domain wall oscillates and radiates gravitational energy and, if one neglects other contributions, the 'equation of motion' is:
\begin{equation}\label{gravo}
	\frac{d \sigma_s }{dt}=-P_{grav}=-G \sigma_s^2,
\end{equation}
where $G$ is the gravitational constant and $\sigma_s$ is the surface energy density of the domain wall.
According to the values reported in Ref.~\cite{Sikivie2008} and the dynamics described by Eq.~(\ref{gravo}), we expect a life-time $\tau_{grav}$:
\begin{equation}\label{taugr}
	\tau_{grav} \sim (G \sigma_s)^{-1} \sim 2 \times 10^6 \, \SI{}{s} \left(\frac{10^{9} \, \SI{}{GeV}}{f_a}\right),
\end{equation}
and a pressure $P_{grav}$:
\begin{equation}
	P_{grav} \sim G \sigma^2_s \sim 0.8375 \left(\frac{f_a}{10^9 \, \SI{}{GeV}}\right)^2 \SI{}{MeV}^4	.
\end{equation}
This last pressure is much bigger than $P_{em}^0$ for every reasonable axion mass, and with a temperature $T \simeq \Lambda_{QCD}$ it is much worse for an actual domain wall since, according to Refs.\cite{Sikivie2008,PhysRevD.30.712}, two axion walls repulse each other with acceleration $a= 2 \pi G \sigma_s$, so they reach a speed $c$ into a time $\tau \sim (2 \pi G \sigma_s)^{-1}$. That means that, if we qualitatively estimate the electromagnetic radiation by considering approximately the velocity-dependent pressure (gotten with zero acceleration)  we get  fainter pressures that approach to zero when $v \rightarrow 1$.   
We can analogously get a partial life-time associated to the electromagnetic  radiation pressure (\ref{Pem0}) that is
\begin{equation}\label{taurad}
\tau_{a \gamma \gamma}=\frac{\sigma_s}{P^0_{em}} \sim 35.75 \times 10^{24}\left(\frac{\Theta_0}{10^{-3}}\right)^{-3}   \left(\frac{m_a}{10^{-2} \, \SI{}{eV}}\right)^{-3}  \left(\frac{T}{ \SI{1}{GeV}}\right)^{-1} \left[1+\ln{\left(10^{-14}\frac{T}{m_a} \right)}   \right]^{-1}\, \SI{}{s} .
\end{equation}
This value is useful for comparisons with what one gets for $N_{DW}>1$ scenarios.
In fact, as mentioned before, the cosmological scenario is problematic for $N_{DW} >1$ where there is the possibility of a wall-dominated Universe with an energy density inconsistent with observations.
As shown e.g. by Ref.~\cite{Sikivie2008}, the typical times for such disaster events would be the times $\tau_2$, where domain wall form,   and the age of the Universe $\tau_{DW}$ where domain walls would dominate when $N_{DW}>1$.

The comparison of the  times $\tau_{DW}$ and $\tau_2$ in the former reference with the very high decay life-time in Eq.~(\ref{taurad}), that is far bigger than the current age of the Universe, suggests us that the electromagnetic radiation pressure is irrelevant for the evolution of cosmological axion domain walls.
\section{Comparison with Faraday Effect and chiral media and about the optical activity in Axion Electrodynamics}\label{optical}
It is worth mentioning that the mode splittings     are analogous to those encountered in the Faraday effect, i.e., the polarization rotation that is  proportional to the longitudinal strong magnetic field. For treatises of this effect, cf., for instance, \cite{prati2003propagation, mansuripur1999faraday}.
Moreover, there is also a strong connection with the Casimir polarization rotation observed in a chiral medium when there is a strong transverse magnetic field present. Cf.
\cite{PhysRevB.99.125403, hoye2020casimir}.
This effect has been treated for Axion Electrodynamics, however with fixed axion field that holds constant space and time derivatives, in Refs. \cite{RevModPhys.93.015004} and \cite{PhysRevD.41.1231}. 
It is indeed easy to notice how our theory with fixed axion background is equivalent to a modified Electrodynamics with an additional Chern-Simons term:
\begin{equation}\label{carroll}
	\mathcal{L}_a=-\frac{1}{2} p_{\alpha} A_{\beta} \tilde{F}^{\alpha \beta},
\end{equation}
if $p_{\alpha}=g_{a \gamma \gamma}\partial_{\alpha} a(x)$, regardless if axion derivatives are constant and uniform or not.\\
The theory of this modified electrodynamics when $p_{\alpha}$ are constants is very well treated in Ref.~\cite{PhysRevD.41.1231}.
Here we mention the similarities and the differences between our results and the ones in \cite{PhysRevD.41.1231,RevModPhys.93.015004}, in order to highlight what we get when considering more physically interesting axion configurations, such as a coherent oscillating axion field and an axion domain wall. 
As found in Refs. \cite{PhysRevD.41.1231,RevModPhys.93.015004}, an Axion Electrodynamics theory with fixed axion field that holds constant and uniform derivatives is characterized by having an optically active vacuum
\begin{equation}
	\frac{d \Phi}{dt}=\frac{1}{2} g_{a \gamma \gamma}  \sqrt{\frac{\mu}{\varepsilon}}\left(\dot{a}+\frac{\omega}{k^2}  \vec{k} \cdot \nabla a    \right).
\end{equation}
This comes from the dispersion relation:
\begin{equation}
\omega_{\pm}=\frac{|\vec{k}|}{\sqrt{\varepsilon \mu}} \pm \frac{1}{2} g_{a \gamma \gamma}  \sqrt{\frac{\mu}{\varepsilon}}\left(\dot{a}+\frac{\omega}{|\vec{k}|^2}  \vec{k} \cdot \nabla a \right)+ \mathcal{O}(g_{a \gamma \gamma}^2),
\end{equation}
which is compatible with the dispersion relations obtained in Subsections (\ref{constdot}) and (\ref{Casimirdisperdo}). However, this is not enough to grasp all the properties of real axion backgrounds, such as an oscillating one and an axion domain wall. These two are characterized by a typical frequency (in the case $a(t)=a_0 \sin(\omega_a t)$ it is the axion frequency $\omega_a$, while for the toy model of domain wall it is $m_L=\frac{\Theta_0}{L}$). The result   is valid in the limit of frequencies and wave numbers much bigger than the typical frequency, because they correspond to modes with very low wavelengths or periods, so they do not 'appreciate' the space-time variations of the derivatives of axion field.
It is worth noticing that the deviations we got for the two systems: a domain wall is characterized by having a more significant reflection coefficient   at  $k_z=m_L=\frac{\Theta_0}{L}$, which means a very steady variation of the polarization plane near to the interface, if compared to the variation of the axion field along the light trail (that is arbitrarily small for an arbitrary small neighborhood of the interface). Similarly, it happens with $a(t)=a_0 \sin(\omega_a t)$ that, in the forementioned phenomenon of axion echo , we have production of a fainter radiation when inputting an electromagnetic wave with frequency $\omega=\frac{\omega_a}{2}$, that lets the polarization plane rotate with a behaviour not described by the theory with constant axion derivatives. Furthermore, when generating a strong magnetic field there is production of faint photons with frequency $\omega=\omega_a$ inside an electromagnetic cavity (see Refs. \cite{RevModPhys.93.015004,Millar_2017,PhysRevD.107.043522}), so there is a phenomenon more similar to the Faraday effect.
Those arguments show how the validity of 'achromaticity of the optical activity of Axion Electrodynamics' is only valid in the regime of 'high frequencies'.
\section{Conclusions}
In this article we have treated the theory of zero-point energy modifications in Axion Electrodynamics with a focus on physical systems of interest for the High Energy Physics axion and the effective axion field in topological insulators. We reviewed the theory of Axion Electrodynamics and particularly the energy-momentum conservation in a linear dielectric and magnetic material.
Adopting the model of the oscillating axion background, we discussed the axion-induced modifications to the Casimir force between two parallel plates by using the Green's function method and suggested a way of enhancing axion-induced Casimir force through a closed string of dielectric point defects. We calculated  the  radiation pressure  acting on a axion domain wall at temperature $T$.
Our results for an oscillating axion field and a domain wall are useful for  "axionic topological insulators" and experimental systems analogous to  the ones showed e.g. by Jiang, Q. D., \& Wilczek, F. (2019) \cite{PhysRevB.99.125403} and Fukushima et al. (2019) \cite{PhysRevD.100.045013}. Finally, we compared our results, where we assume time-dependent or space-dependent axion configurations, with the discussion of the optical activity of Axion Electrodynamics by Sikivie (2021) \cite{RevModPhys.93.015004} and Carrol et al. (1990) \cite{PhysRevD.41.1231} and showed the 'cromaticity' of the vacuum in Axion Electrodynamics.

	\textbf{Note:} During the revision of our paper, a recent work ~\cite{ema2023zero} has appeared on arXiv, where the Authors discuss also finite-temperature Casimir force in Axion Electrodynamics for other cases of interest.

\section*{Acknowledgments}
We thank the Erasmus project, the Norwegian University of Science and Technology, and the University of Palermo, for the support of the Erasmus Traineeship for Amedeo Maria Favitta. Part of  this work was performed in connection with his  Master Thesis.
We are grateful  to  Yuri N. Obukhov,  Markku Oksanen  and Roberto Passante for valuable informations and several suggestions.

\bibliographystyle{apsrev4-2}
\bibliography{mastertesi}
\end{document}